\documentclass[11pt,a4paper]{article}
\pdfoutput=1
\usepackage{jheppub}
\usepackage{amsmath}
\usepackage{amssymb}
\usepackage{graphics}
\usepackage[active]{srcltx}
\usepackage{pdfsync}
\usepackage{shuffle}
\usepackage{slashed}

\usepackage{subfigure}

\newcommand{\bea}{\begin{eqnarray}}
\newcommand{\eea}{\end{eqnarray}}
\newcommand{\nn}{\nonumber \\}

\def\W #1{\widetilde{#1}}

\def\Tr{\mathop{\rm Tr}}

\def\eref#1{(\ref{#1})}

\def\a{{\alpha}}

\def\b{{\beta}}

\title{New recursive construction for tree NLSM and SG amplitudes, and new understanding of enhanced Adler zero}
\author[a]{Kang Zhou,}

\affiliation[a]{Center for Gravitation and Cosmology, College of Physical Science and Technology, Yangzhou University,\\
No.180, Siwangting Road, Yangzhou, 225009, P.R. China.}

\emailAdd{zhoukang@yzu.edu.cn}

\date{\today}
\abstract{ We propose a new bottom up method to construct tree amplitudes of non-linear sigma model (NLSM) and special Galileon theory (SG), based on assuming the universality of soft behaviors and the double copy structure. We extend the on-shell amplitudes to off-shell ones with two off-shell external legs, which allow the numbers of external legs to be odd. Then the $3$-point and $4$-point off-shell amplitudes can be bootstrapped, and the soft behaviors of $4$-point NLSM and SG amplitudes can be derived from them. The universality of soft behaviors allows us to invert the resulted soft theorems to construct higher-point off-shell amplitudes recursively, and express them in the formula of expansions to tree amplitudes of bi-adjoint scalar theory. We emphasize that the exact forms of universal soft behaviors are derived, rather than assumed as the input. Back to the on-shell limit, amplitudes with odd numbers of external legs vanish automatically, and the enhanced Adler zero emerge. From the bottom up perspective without the aid of a Lagrangian, the enhanced Adler zero are understood as that soft behaviors vanish faster than the degree expected from the naive power counting of soft momentum in the formula of expansions. Interestingly, such "zero" have explicit formulas and can be interpreted naturally. For tree amplitudes of Born-Infeld and Dirac-Born-Infeld theories, our method for construction does not make sense, but the enhanced Adler zero can be studied similarly.
}

\begin{document}

\maketitle \flushbottom

\section{Introduction}
\label{sec-intro}

One crucial aim of modern S-matrix program is to construct scattering amplitudes directly by exploiting general principles like Lorentz invariance and unitarity, with out the aid of a traditional Lagrangian. A well known example is that a wide range of tree amplitudes are constructible via Britto-Cachazo-Feng-Witten (BCFW) on-shell recursion which naturally takes factorization as the tool \cite{Britto:2004ap,Britto:2005fq}. Subsequently, BCFW recursion relation led to amazing progress in reformulations of perturbative quantum field theory, for instance the positive Grassmannian formula for planer ${\cal N}=4$ super Yang-Mills theory, and Amplituhedron \cite{Arkani-Hamed:2012zlh,Arkani-Hamed:2013jha,Arkani-Hamed:2013kca}. Meanwhile, another type of bottom up construction, based on soft theorems, have also caught researches' attentions. Soft theorems were first discovered at the leading order, for photons and gravitons \cite{Low,Weinberg}. In 2014, they have been revived for tree gravitational (GR) and Yang-Mills (YM) amplitudes at higher orders \cite{Cachazo:2014fwa,Casali:2014xpa}, and were generalized to arbitrary space-time dimensions \cite{Schwab:2014xua,Afkhami-Jeddi:2014fia}. The soft theorems were widely used in construction of tree amplitudes, such as the inverse soft theorem program, and using another type of soft behavior called the Adler zero to construct amplitudes of a variety of effective theories \cite{Cheung:2014dqa,Cheung:2015ota,Cheung:2016drk,Luo:2015tat,Cheung:2018oki,Elvang:2018dco,Cachazo:2016njl,Rodina:2018pcb,Boucher-Veronneau:2011rwd,Nguyen:2009jk}.

The constructions grounded on soft theorems mentioned above requires the knowledge of exact soft behaviors, such as the explicit formulas of soft factors, or the assumption of Adler zero. From the bottom up perspective, the exact soft behaviors serve as the input, thus require the interpretation for the origin of them without using any top down derivation. In the recent work, we made an attempt to bypass this disadvantage \cite{Zhou:2022orv}. We solely assumed the universality of soft behaviors, as well as the double copy structure of tree amplitudes \cite{Kawai:1985xq,Bern:2008qj,Chiodaroli:2014xia,Johansson:2015oia,Johansson:2019dnu,Cachazo:2013iea}. It was found that the tree Yang-Mills-scalar (YMS) and pure Yang-Mills (YM) amplitudes, as well as the soft factors themselves, can be completely determined by imposing above two assumptions. The resulted YMS and YM amplitudes are expressed in the formula of expanding them into amplitudes of bi-adjoint scalars (BAS) theory. The double copy structure indicates that these expansions can be extended to Einstein-Yang-Mills (EYM) and gravity (GR) theories directly.

Then a natural question arisen, can the method in \cite{Zhou:2022orv} be applied to other theories? Obviously, it is equivalent to ask which amplitudes have universal soft behaviors and satisfy the double copy structure. From the perspective of Cachazo-He-Yuan (CHY) formalism \cite{Cachazo:2013gna,Cachazo:2013hca, Cachazo:2013iea, Cachazo:2014nsa,Cachazo:2014xea}, the most natural candidates are effective theories including non-linear sigma model (NLSM), special  Galileon theory (SG), Born-Infeld theory (BI), as well as Dirac-Born-Infeld theory (DBI). In CHY framework, tree amplitudes of these theories can be generated from GR ones, through the so called compactifying, squeezing and the generalized dimensional reduction procedures \cite{Cachazo:2014xea}, or acting appropriate differential operators \cite{Cheung:2017ems,Zhou:2018wvn,Bollmann:2018edb}. The resulted amplitudes satisfy double copy structure automatically, and it is natural to expect them to inherit the universality of soft behavior of GR amplitudes.

Motivated by the above consideration, in \cite{Zhou:2023quv} we bootstrapped the tree NLSM amplitudes by imposing only universality of soft behavior and the double copy structure. However, the obtained result does not fully satisfy our expectation. We found the correct expansion of NLSM amplitude to BAS ones arises as a natural solution of imposed constraints. However, since the independence of BAS amplitudes indicated by Bern-Carrasco-Johansson (BCJ) relations \cite{Bern:2008qj,Chiodaroli:2014xia,Johansson:2015oia,Johansson:2019dnu}, it is hard to ensure the uniqueness of the solution.

This paper is the continuation of the previous investigation in \cite{Zhou:2023quv}. Our purpose is to develop a bottom up method based on two assumptions which are the universality of soft behaviors and the double copy structure. The method should allow us to construct higher-point amplitudes recursively from the lowest-point ones which can be fixed by bootstrapping. In particular, the role of exact soft behaviors should not be the input. To realize the goal, it seems that one need to find the universal double soft behavior by considering the $6$-point and $4$-point amplitudes, then use the resulted double soft behavior to bootstrap higher-point amplitudes, since the un-vanishing on-shell NLSM amplitudes require even numbers of external legs. However, while the $4$-point amplitudes can be completely determined by bootstrapping, the derivation of $6$-point ones requires the using of other method. Thus, to make our construction to be self-contained, the idea described above does not make sense.

To solve this difficulty, we extend the on-shell tree NLSM amplitudes to off-shell ones by introducing two off-shell external legs. Such extension allows the existence of un-vanishing amplitudes with odd numbers of external legs. In the off-shell case, both $3$-point and $4$-point amplitudes can be uniquely fixed by bootstrapping based one some natural physical conditions. Then, one can derive the single soft behavior by analysing $3$-point and $4$-point amplitudes, and impose the universality of resulted soft behavior to construct higher-point ones recursively. The obtained general amplitudes are expressed in the formula of expanding them to corresponding off-shell BAS amplitudes. In the on-shell limit, the off-shell amplitudes with odd number of external legs vanish automatically.

The universal soft behavior tend to zero in the on-shell limit, this observation leads to the new understanding of enhanced Adler zero from the bottom up perspective, without the aid of a Lagrangian and associated symmetries. The enhanced Adler zero is the phenomenon that amplitudes have vanished soft behavior which exceed the degree expected from naive derivative power counting in Lagrangian \cite{Cheung:2014dqa,Cheung:2015ota,Cheung:2016drk}. Interestingly, the NLSM, SG and DBI theories under consideration in this paper are the only "exceptional" effective theories for pure scalars satisfying the enhanced Adler zero, lie on the boundary of allowed theory space \cite{Cheung:2016drk}. From our bottom up point of view, the enhanced Adler zero should be understood as that amplitudes vanish faster than the degree expected from the power counting of soft momentum in the formula of expansion. Our result shows that,
such "zero" has special explicit formula, for the NLSM case it is governed by a simple identity \eref{iden} which will be introduced in latter sections, as well as the vanishing of amplitudes with odd number of external legs.

The paralleled manipulation can be applied to construct tree SG amplitudes, and the enhanced Adler zero for SG amplitudes can be understood similarly. As will be seen, the vanishing of tree SG amplitudes at the $\tau^1$ order is interpreted by the simple relation $\sum_{\ell\neq j}k_j\cdot k_\ell=0$ due to the momentum conservation and on-shell condition, as well as the vanishing of amplitudes with odd number of legs. Here $\tau$ is the parameter for labeling different orders of soft behavior, namely, we re-scale the momentum $k_i$ carried by the external leg $i$ as $k_i\to\tau k_i$, and expand the amplitude in $\tau$. The vanishing of SG amplitudes at the $\tau^2$ order can also be interpreted straightforwardly. For tree BI and DBI amplitudes, our bottom up construction does not make sense, due to the reasons which will be explained in section.\ref{sec-expan-toNLSM}. However, after deriving the expansions of these amplitudes to BAS ones by employing other methods, the associated enhanced Adler zero can be understood in the similar way. In particular, the Adler zero for BI amplitudes can be interpreted by the simple relation
$\sum_{\ell\neq j}\epsilon_j\cdot k_\ell=0$, and the vanishing of amplitudes with odd number of external legs.

The remainder of this paper is organized as follows. In section.\ref{sec-background}, we rapidly review the necessary background, including the double color ordered tree BAS amplitudes and the corresponding soft behavior at the leading order, as well as the expansions of other amplitudes to those BAS ones. In section.\ref{sec-expan}, we propose the new recursive method to construct the tree NLSM amplitudes recursively, and understand the enhanced Adler zero of them via the purely bottom up point of view. In section.\ref{sec-SG}, we apply the same idea to construct the tree SG amplitudes and interpret their enhanced Adler zero. In section.\ref{sec-expan-toNLSM}, we study the enhanced Adler zero for tree BI and DBI amplitudes along the similar line. In section.\ref{sec-conclu}, we end with a brief summery and discussion.

\section{Background}
\label{sec-background}

In this section we give a brief review for the necessary background. In subsection.\ref{subsecBAS}, we introduce the tree level amplitudes of bi-adjoint scalar (BAS) theory, as well as the corresponding soft behavior at the leading order. Some notations and conventions which will be used subsequently are also included. In subsection.\ref{subsecexpand}, we rapidly discuss the expansions of tree amplitudes to BAS amplitudes, including the existence of such expansions, the choice of basis, and the characters of coefficients.

\subsection{Tree level BAS amplitudes}
\label{subsecBAS}

\begin{figure}
  \centering
  \includegraphics[width=6cm]{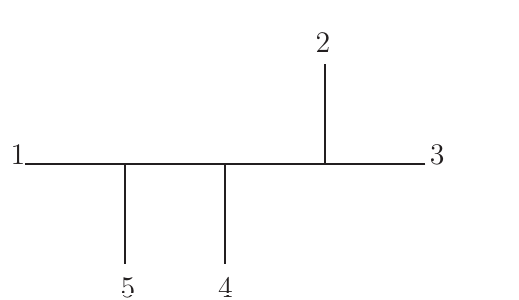}
   \includegraphics[width=6cm]{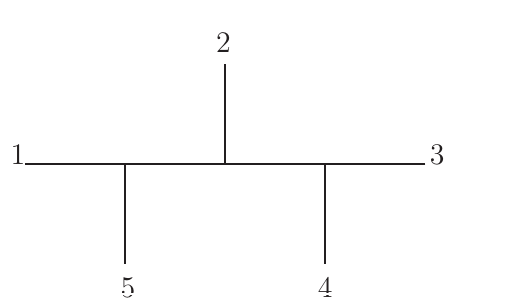}  \\
  \caption{Two $5$-point diagrams}\label{5p}
\end{figure}

The bi-adjoint-scalar (BAS) theory is the theory for massless scalar fields $\phi_{a\bar{a}}$ with the Lagrangian
\bea
{\cal L}_{\rm BAS}={1\over2}\,\partial_\mu\phi^{a\bar{a}}\,\partial^{\mu}\phi^{a\bar{a}}-{\lambda\over3!}\,f^{abc}f^{\bar{a}\bar{b}\bar{c}}\,
\phi^{a\bar{a}}\phi^{b\bar{b}}\phi^{c\bar{c}}\,,~~\label{Lag-BAS}
\eea
where the structure constant $f^{abc}$ and generator $T^a$ are related by
\bea
[T^a,T^b]=if^{abc}T^c\,,
\eea
and the dual color algebra encoded by $f^{\bar{a}\bar{b}\bar{c}}$ and $T^{\bar{a}}$ is analogous.
The tree level amplitudes of this theory contain only propagators, without any numerator, and can be decomposed into partial amplitudes with coefficients $\Tr(T^{a_{\sigma_1}}\cdots T^{a_{\sigma_n}})\cdot\Tr(T^{\bar{a}_{\bar{\sigma}_1}}\cdots T^{\bar{a}_{\bar{\sigma}_n}})$,
where $\sigma_i$ and $\bar{\sigma}_i$ denote permutations among all external scalars.
Each partial amplitude is double color ordered, i.e., it exhibits planarity with respect to two color orderings simultaneously \cite{Cachazo:2013iea}. We take the $5$-point amplitude ${\cal A}_{\rm BAS}(1,2,3,4,5|1,4,2,3,5)$ as an example.
In Figure.\ref{5p}, the first planer Feynman diagram satisfies both two orderings $(1,2,3,4,5)$ and $(1,4,2,3,5)$, while the second one violates the ordering
$(1,4,2,3,5)$. Thus, two orderings $(1,2,3,4,5|1,4,2,3,5)$ permits the first diagram, and forbids the second. One can draw all planer diagrams correspond to the ordering $(1,2,3,4,5)$, and show that the first diagram in Figure.\ref{5p} is the only candidate allowed by $(1,4,2,3,5)$. Thus, the tree BAS amplitude ${\cal A}_{\rm BAS}(1,2,3,4,5|1,4,2,3,5)$ can be computed as
\bea
{\cal A}_{\rm BAS}(1,2,3,4,5|1,4,2,3,5)={1\over s_{23}}{1\over s_{51}}\,,
\eea
up to an overall sign. The Mandelstam variable $s_{i\cdots j}$ is defined as
\bea
s_{i\cdots j}\equiv k_{i\cdots j}^2\,,~~~~k_{i\cdots j}\equiv\sum_{a=i}^j\,k_a\,,~~~~\label{mandelstam}
\eea
where $k_a$ is the momentum carried by the external leg $a$.

In general, each double color ordered partial amplitudes carry an overall sign $\pm$, since the anti-symmetry of structure constants $f^{abc}$ indicates that swapping two lines at a common vertex creates a $-$. In this paper we chose the overall sign to be $+$ if two orderings carried by the partial BAS amplitude are the same. This convention is chosen for latter convenience when discussing the soft behavior of BAS scalars, and is different from that in \cite{Cachazo:2013iea}. For instance, the new convention indicates the overall sign $+$ for the amplitude ${\cal A}_{\rm BAS}(1,2,3,4|1,2,3,4)$, while the old one in \cite{Cachazo:2013iea} gives $-$. After fixing the sign for the special case with two same orderings, the sign carried by other partial BAS amplitudes with general orderings can be determined by counting the number of flipping. Thus, the relative $\pm$ among different partial BAS amplitudes have not been altered when comparing with rules in \cite{Cachazo:2013iea}.

Then we discuss the soft behavior of double color ordered tree BAS amplitudes, which is crucial in subsequent sections.
When considering the leading order soft behavior, the $2$-point channels play the central role. Since the partial BAS amplitude carries two color orderings, if the $2$-point channel contributes $1/s_{ab}$ to the amplitude, external legs $a$ and $b$ must be adjacent to each other in both two orderings.
To denote this feature, we introduce the symbol $\delta_{ab}$\footnote{The Kronecker symbol will not appear in this paper, thus we hope the notation $\delta_{ab}$ will not confuse the readers.}, defined as follows \cite{Zhou:2022orv}. In an ordering $\sigma_n$, if two legs $a$ and $b$ are adjacent, then $\delta_{ab}=1$ if $a$ precedes $b$, and $\delta_{ab}=-1$ if $a$ follows $b$. If $a$ and $b$ are not adjacent, $\delta_{ab}=0$. From the definition, it is straightforward to observe $\delta_{ab}=-\delta_{ba}$, as well as a simple but useful identity
\bea
\sum_{b\neq a}\,\delta_{ab}=0\,.~~\label{iden}
\eea

Using the symbol $\delta_{ab}$, we now give the explicit formulas of soft behavior of partial BAS amplitudes at the leading order. For the double color ordered BAS amplitude ${\cal A}_{\rm BAS}(1,\cdots,n|\sigma_n)$, we re-scale $k_i$
as $k_i\to\tau k_i$, and expand the amplitude in $\tau$. The leading order contribution manifestly aries from $2$-point channels $1/s_{1(i+1)}$
and $1/ s_{(i-1)i}$ which provide the $1/\tau$ order contributions, namely,
\bea
{\cal A}^{(0)_i}_{\rm BAS}(1,\cdots,n|\sigma_n)&=&{1\over \tau}\Big({\delta_{i(i+1)}\over s_{i(i+1)}}+{\delta_{(i-1)i}\over s_{(i-1)i}}\Big)\,
{\cal A}_{\rm BAS}(1,\cdots,i-1,\not{i},i+1,\cdots,n|\sigma_n\setminus i)\nn
&=&S^{(0)_i}_{\rm BAS}\,{\cal A}_{\rm BAS}(1,\cdots,i-1,i+1,\cdots,n|\sigma_n\setminus i)\,,~~~\label{soft-theo-s}
\eea
where $\not{i}$ means removing the leg $i$, $\sigma_n\setminus i$ denotes the color ordering generated from $\sigma_n$ by eliminating $i$, and $\delta_{ab}$ are defined from the ordering $\sigma_n$. We have introduced the superscript $(0)_i$ to denote the leading order contribution when $k_i\to\tau k_i$. The leading soft factor $S^{(0)_i}_{\rm BAS}$ for the scalar $i$ is extracted as
\bea
S^{(0)_i}_{\rm BAS}={1\over \tau}\,\Big({\delta_{i(i+1)}\over s_{i(i+1)}}+{\delta_{(i-1)i}\over s_{(i-1)i}}\Big)\,.~~~~\label{soft-fac-s-0}
\eea
We emphasize that from \eref{soft-theo-s} one can see that the definition of $\delta_{ab}$ is coincide with our convention for the overall $\pm$ sign. Suppose two orderings of the BAS amplitude are the same, namely $\sigma_{n}=(1,\cdots,n)$, then $\delta_{(i-1)i}=\delta_{i(i+1)}=1$,
and \eref{soft-theo-s} carries the correct $\pm$ sign since both ${\cal A}_{\rm BAS}(1,\cdots,n|1,\cdots,n)$ and ${\cal A}_{\rm BAS}(1,\cdots,i-1,i+1,\cdots,n|1,\cdots,i-1,i+1,\cdots,n)$ carry $+$. If one exchange legs $i-1$ and $i$ (or $i$ and $i+1$) in $\sigma_n$, the relative $-$ is absorbed into $\delta_{(i-1)i}$ (or $\delta_{i(i+1)}$), since $\delta_{ab}=-\delta_{ba}$. If one perform other flipping in $\sigma_n$, the $\pm$ sign carried by ${\cal A}_{\rm BAS}(1,\cdots,n|\sigma_n)$ and ${\cal A}_{\rm BAS}(1,\cdots,i-1,i+1,\cdots,n|\sigma_n\setminus i)$ changes simultaneously, thus \eref{soft-theo-s} still holds.

Another equivalent expression is
\bea
{\cal A}^{(0)_i}_{\rm BAS}(\sigma_n|\W\sigma_n)
&=&S^{(0)_i}_{\rm BAS}\,{\cal A}_{\rm BAS}(\sigma_n\setminus i|\W\sigma_n\setminus i)\,,~~~\label{soft-theo-s-2}
\eea
where
\bea
S^{(0)_i}_{\rm BAS}={1\over \tau}\,\sum_{j\in\{1,\cdots,n\}\setminus i}\,{\delta_{ij}\,\W\delta_{ij}\over s_{ij}}\,.~~~~\label{soft-fac-s-2}
\eea
Here $\delta_{ab}$ is defined for the ordering $\sigma_n$, while $\W\delta_{ab}$ is defined for $\W\sigma_{ab}$.
The equivalence between two expressions \eref{soft-theo-s} and \eref{soft-theo-s-2} can be verified directly via the definition of $\delta_{ab}$.
Both two expressions will be used in subsequent sections.

In this paper, we will also consider the extension of on-shell BAS amplitudes which turns external momenta $k_1$ and $k_n$ to be off-shell.
When $k_1^2\neq0$, $k_n^2\neq0$, the Mandelstam variables are modified as
\bea
s_{1i}=2k_1\cdot k_i\,&\to&\,s_{1i}=k_1^2+2k_1\cdot k_i\,,\nn
s_{in}=2k_i\cdot k_n\,&\to&\,s_{in}=k_n^2+2k_i\cdot k_n\,,\nn
s_{1n}=2k_1\cdot k_n\,&\to&\,s_{1n}=k_1^2+k_n^2+2k_1\cdot k_n\,,
\eea
while $s_{ij}$ with $i,j\in\{2,\cdots,n-1\}$ remains un-altered. For such off-shell case, propagators $1/s_{1i}$ and $1/s_{in}$ no longer contribute $1/\tau$ when $k_i\to\tau k_i$, thus one need to remove $\delta_{1i}/s_{1i}$ and $\delta_{in}/s_{in}$ from the soft factor in \eref{soft-fac-s-0}. The more formal expression can be generated from \eref{soft-fac-s-2} by removing contributions from $j=1,n$, namely,
\bea
S^{(0)_i}_{\rm BAS}={1\over \tau}\,\sum_{j\in\{2,\cdots,n-1\}\setminus i}\,{\delta_{ij}\,\W\delta_{ij}\over s_{ij}}\,.~~~~\label{soft-fac-s-2-offshell}
\eea
%

\subsection{Expanding tree level amplitudes to BAS basis}
\label{subsecexpand}

\begin{figure}
  \centering
   \includegraphics[width=8cm]{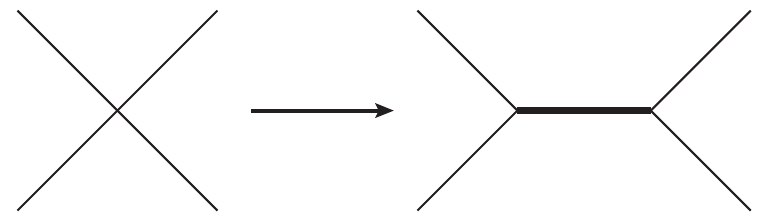}\\
  \caption{Turn the $4$-point vertex to $3$-point ones. The bold line corresponds to the inserted propagator $1/D$. This manipulation turns the original numerator $N$ to $DN$, and split the original coupling constant $g$ to two $\sqrt{g}$ for two cubic vertices.}\label{3pto4p}
\end{figure}

Tree level amplitudes for massless particles and cubic interactions can be expanded to double color ordered BAS amplitudes,
due to the observation that any Feynman diagram contributes to at least one partial BAS amplitude. For higher-point vertices, one can decompose them into cubic ones by inserting $1=D/D$ with the propagator $1/D$ and the numerator $D$, an example is shown in Figure.\ref{3pto4p}. Based on such insertions, one can decompose each tree amplitude to tree Feynman diagrams with only cubic interactions. Since each Feynman diagram contributes propagators which can be provided by BAS amplitudes, along with a numerator depends on kinematical variables, one can conclude that each tree amplitude for massless particles can be expanded to double color ordered partial BAS amplitudes, with coefficients which are polynomials depend on Lorentz invariants created by external kinematical variables.

The appropriate basis for expansion can be determined by employing the well known Kleiss-Kuijf (KK) relation \cite{Kleiss:1988ne}
\bea
{\cal A}_{\rm BAS}(1,\vec{\pmb{\a}},n,\vec{\pmb{\b}}|\sigma_n)=(-)^{|\pmb{\b}|}\,{\cal A}_{\rm BAS}(1,\vec{\pmb{\a}}\shuffle\vec{\pmb{\b}}^T,n|\sigma_n)\,,~~~\label{KK}
\eea
where $\vec{\pmb{\a}}$ and $\vec{\pmb{\b}}$ are two ordered subsets of external scalars, and $\vec{\pmb{\b}}^T$ is the ordered set generated from $\vec{\pmb{\b}}$ by reversing the original ordering. The notation ${|\pmb{\b}|}$ stands for the length of the set $\vec{\pmb{\b}}$, i.e., the number of elements in $\vec{\pmb{\b}}$. The $n$-point BAS amplitude ${\cal A}_{\rm BAS}(1,\vec{\pmb{\a}},n,\vec{\pmb{\b}}|\sigma_n)$ at the l.h.s of \eref{KK} carries two color orderings, one is $(1,\vec{\pmb{\a}},n,\vec{\pmb{\b}})$, another one is encoded by $\sigma_n$. The symbol $\shuffle$ means summing over all
possible shuffles of two ordered sets $\vec{\pmb{\b}}_1$ and $\vec{\pmb{\b}}_2$, i.e., all permutations in the set $\vec{\pmb{\b}}_1\cup \vec{\pmb{\b}}_2$ while preserving the orderings
of $\vec{\pmb{\b}}_1$ and $\vec{\pmb{\b}}_2$. For instance, suppose $\vec{\pmb{\b}}_1=\{1,2\}$ and $\vec{\pmb{\b}}_2=\{3,4\}$, then
\bea
{\cal A}(\vec{\pmb{\b}}_1\shuffle \vec{\pmb{\b}}_2)&=&{\cal A}(1,2,3,4)+{\cal A}(1,3,2,4)+{\cal A}(1,3,4,2)\nn
& &+{\cal A}(3,1,2,4)+{\cal A}(3,1,4,2)+{\cal A}(3,4,1,2)\,.~~~~\label{shuffle}
\eea
The analogous KK relation holds for another color ordering $\sigma_n$.
The KK relation implies the independence of double color ordered partial BAS amplitudes, thus the basis can be chosen as BAS amplitudes
${\cal A}_{\rm BAS}(1,\sigma_1,n|1,\sigma_2,n)$, with $1$ and $n$ are fixed at two ends in each ordering. Such basis is called the KK BAS basis. Based on the discussion above, any amplitude for massless particles can be expanded to this basis.
The basis provides propagators, while the coefficients in expansions provide numerators without any pole. In this sense, one can regard the BAS KK basis as the combinations of massless propagators, without knowing the associated Lagrangian in \eref{Lag-BAS}.

As well known, the double color ordered BAS amplitudes are also connected by Bern-Carrasco-Johansson (BCJ) relations \cite{Bern:2008qj,Chiodaroli:2014xia,Johansson:2015oia,Johansson:2019dnu}. Here we give the form of the fundamental BCJ relation,
\bea
\prod_{i=2}^{n-1}\,\Big(\sum_{j=1}^{i-1}\,s_{ij}\Big)\,{\cal A}_{\rm BAS}(1,\cdots,n|\sigma_n)=0\,,~~\label{bcj}
\eea
since it will be used later.
Constrained by BCJ relations, the independent BAS amplitudes are those three legs are fixed at three particular positions in the color orderings. However, in BCJ relations, coefficients of BAS amplitudes depend on kinematical variables, this character leads to expansions in which the coefficients include poles. In this paper, we choose the KK basis since we hope all poles are included in basis, and coefficients only serve as numerators.

In subsequent sections, we will consider expansions of NLSM, SG, BI and DBI amplitudes. The color ordered tree NLSM amplitude ${\cal A}_{\rm NLS}(1,\sigma_{n-2},n)$ can be expanded to KK BAS basis as
\bea
{\cal A}_{\rm NLS}(1,\W\sigma_{n-2},n)=\sum_{\sigma_{n-2}}\,{\cal C}(\sigma_{n-2},k_i)\,{\cal A}_{\rm BAS}(1,\sigma_{n-2},n|1,\W\sigma_{n-2},n)\,,~~\label{toKK-1}
\eea
where $\sigma_{n-2}$ and $\W\sigma_{n-2}$ are orderings among $(n-2)$ external legs in $\{2,\cdots,n-1\}$. The double copy structure \cite{Kawai:1985xq,Bern:2008qj,Chiodaroli:2014xia,Johansson:2015oia,Johansson:2019dnu,Cachazo:2013iea} indicates that the coefficient ${\cal C}(\sigma_{n-2},k_i)$ depends on
momenta $k_i$ carried by external scalars, orderings $\sigma_{n-2}$, but is independent of the ordering $\W\sigma_{n-2}$\footnote{The standard double copy means the GR amplitude can be factorized as ${\cal A}_{\rm G}={\cal A}_{\rm YM}\times {\cal S}\times{\cal A}_{\rm YM}$, where the kernel ${\cal S}$ is the inverse of BAS amplitudes. The assumption that the coefficients depend on only one color ordering is equivalent to the standard description, see in \cite{Zhou:2022orv}.}. In \eref{toKK-1}, the ordering of the NLSM amplitude is chosen to satisfy the requirement of KK basis. Since ${\cal C}(\sigma_{n-2},k_i)$ is independent of $\W\sigma_{n-2}$, one can replace $(1,\W\sigma_{n-2},n)$ by the general ordering $\sigma_n$ without fixing any leg at any position, to obtain the ansatz
\bea
{\cal A}_{\rm NLS}(\sigma_n)=\sum_{\sigma_{n-2}}\,{\cal C}(\sigma_{n-2},k_i)\,{\cal A}_{\rm BAS}(1,\sigma_{n-2},n|\sigma_n)\,.~~~\label{exp-N-KK}
\eea
The construction for NLSM amplitudes ${\cal A}_{\rm NLS}(\sigma_n)$ is shifted to the construction for coefficients ${\cal C}(\sigma_{n-2},k_i)$.
We emphasize that the existence of the expansion to BAS KK basis can be proven, but the double copy structure is an assumption.

Tree SG, BI and DBI amplitudes can be double expanded to BAS KK basis as
\bea
{\cal A}(\{i\}_n)=\sum_{\sigma_{n-2}}\,\sum_{\W\sigma_{n-2}}\,{\cal C}(\sigma_{n-2},k_i)\,\W{\cal C}(\W\sigma_{n-2},k_i)\,{\cal A}_{\rm BAS}(1,\sigma_{n-2},n|1,\W\sigma_{n-2},n)\,,~~\label{expan-nocolor}
\eea
where coefficients ${\cal C}(\sigma_{n-2},k_i)$ and $\W{\cal C}(\W\sigma_{n-2},k_i)$ satisfy the double copy structure. Here $\{i\}_n$ denotes the set of external legs in $\{1,\cdots,n\}$, without any ordering. In other words, we use $\{i\}_n$ to distinguish un-ordered amplitudes from ordered ones. In \eref{expan-nocolor}, we see that both two orderings $\sigma_{n-2}$ and $\W\sigma_{n-2}$ are summed over, coincide with the absence of color ordering at the l.h.s.
For BI amplitudes, ${\cal C}(\sigma_{n-2},k_i)$ or $\W{\cal C}(\W\sigma_{n-2},k_i)$ also depend on polarizations $\epsilon_i$ carried by BI photons.

\section{Tree NLSM amplitudes}
\label{sec-expan}

In this section we construct color ordered tree NLSM amplitudes, by using the method based on the universality of soft behavior.
The standard NLSM lagrangian in the Cayley parametrization is given as
\bea
{\cal L}_{\rm NLS}={1\over 8\lambda^2}\,{\rm Tr}(\partial_\mu {\rm U}^\dag\partial^\mu {\rm U})\,,~~\label{Lag-N}
\eea
with
\bea
{\rm U}=(\mathbb{I}+\lambda\Phi)\,(\mathbb{I}-\lambda\Phi)^{-1}\,,
\eea
where $\mathbb{I}$ is the identity matrix, and $\Phi=\phi_IT^I$, with $T^I$ the generators of $U(N)$. Fields $\phi_I$ describe massless scalars, and the accompanied generators $T^I$ indicates the color ordering for the associated partial tree amplitudes.
In the Lagrangian \eref{Lag-N}, the mass dimension of coupling constant $\lambda$ is $(2-d)/2$, in $d$ dimensional space-time. The mass dimension of any $n$-point amplitude is $d-{d-2\over2}n$, thus the kinematic part have mass dimension $2$ since the coupling constants contribute $(2-d)(n-2)/2$.

Thus, our purpose is to construct the $n$-point tree amplitudes ${\cal A}_{\rm NLS}(\sigma_n)$ (kinematic part without coupling constants) for pure massless scalars, with the color ordering $\sigma_n$ and mass dimension $2$. We also assume that the amplitudes have universal soft behavior, and satisfy the double copy structure.  As will be shown, the candidate satisfies all above requirements is unique. In this sense, one can ignore the corresponding Lagrangian description \eref{Lag-N} in the reminder of this section.

Tree amplitudes can be expressed in various formulas, such as the summation of contributions from Feynman diagrams, the CHY integrals, and so on. In this paper we express tree amplitudes as the expansions of them to KK BAS basis discussed in subsection.\ref{subsecexpand} in section.\ref{sec-background}. As pointed out in subsection.\ref{subsecexpand}, the construction of tree NLSM amplitudes is shifted to the construction of coefficients ${\cal C}(\sigma_{n-2},k_i)$ in \eref{exp-N-KK}. We emphasize again the existence of expansions to KK BAS basis can be proven, while the double copy structure which ensures that coefficients ${\cal C}(\sigma_{n-2},k_i)$ depend on only one color ordering carried by BAS amplitudes is an assumption.
In the kinematic part of the amplitude, the propagators contribute the mass dimension $-2(n-3)$, thus the mass dimension of numerators must be $2(n-2)$.
This is the mass dimension of coefficients ${\cal C}(\sigma_{n-2},k_i)$.

In the reminder of this section, we construct the tree NLSM amplitudes recursively. In subsection.\ref{subsec-34p}, we show that the NLSM amplitudes with lowest number of external particles are $4$-point ones, and fix them by bootstrapping. The NLSM amplitudes with odd number of external legs do not exist, this is the obstacle for applying our recursive method. Thus, insubsection.\ref{subsec-offshell}, we extend on-shell NLSM amplitudes to off-shell ones which do not vanish when the number of external legs is odd. The resulted off-shell $4$-point amplitudes have the un-vanishing leading order soft behavior: the amplitude factorizes as the soft factor and the $3$-point amplitude.
Then, in subsection.\ref{subsec-5p} and subsection.\ref{subsec-np}, we generate the $(m+1)$-point off-shell amplitude from the $m$-point one, via the method based on assuming the universality of such soft behavior. The whole process only uses the following assumptions.
\begin{itemize}
\item The amplitude describes the scattering of massless scalars with single coupling constant.
\item The kinematic part of amplitude has mass dimension $2$.
\item The amplitude is color ordered.
\item The universality of single soft behavior: the single soft behavior of lowest $4$-point amplitudes holds for general higher-point ones.
\item The double copy structure: when expanding to BAS basis, coefficients ${\cal C}(\sigma_{n-2},k_i)$ in \eref{exp-N-KK} are independent of the ordering $\sigma_n$.
\item The manifest permutation symmetry among legs in the set $\{2,\cdots,n-1\}$.
\end{itemize}
The on-shell physical amplitudes serve as the special case of off-shell ones, and the enhanced Adler zero for NLSM amplitudes is interpreted by the vanishing of both the soft factor and the amplitude with the odd number of external legs, as can be seen in subsection.\ref{subsec-np}.

\subsection{$4$-point NLSM amplitudes }
\label{subsec-34p}

The $3$-point tree NLSM amplitude has mass dimension $2$, and does not contain any pole. One can never use three on-shell massless
momenta satisfying momentum conservation to construct any un-vanishing Lorentz invariant with mass dimension $2$. Thus, the $3$-point NLSM amplitude does not exist.

The simplest NLSM amplitudes are the $4$-point ones ${\cal A}_{\rm NLS}(\sigma_4)$. The absence of $3$-point amplitude implies that the $4$-point ones
do not have any pole. Then, the mass dimension requires the $4$-point amplitudes to be linear combinations of Mandelstam variables $s$, $t$ and $u$, where $s=s_{12}$, $t=s_{14}$, $u=s_{13}$, satisfying $s+u+t=0$. Such combinations can be determined by considering the symmetry. For ${\cal A}_{\rm NLS}(1,2,3,4)$, the color ordering indicates the symmetry among $s$ and $t$, thus ${\cal A}_{\rm NLS}(1,2,3,4)$ is proportional to $s+t$ or $u$.
We can choose
\bea
{\cal A}_{\rm NLS}(1,2,3,4)=s+t=-u\,,~~\label{4p1}
\eea
via an overall re-scaling of amplitude. Similarly, we have
\bea
& &{\cal A}_{\rm NLS}(1,3,2,4)=t+u=-s\,,\nn
& &{\cal A}_{\rm NLS}(1,2,4,3)=u+s=-t\,.~~\label{4p2}
\eea

The $4$-point tree NLSM amplitudes can be expanded to KK BAS basis, the double copy assumption requires the following expanded formula
\bea
{\cal A}_{\rm NLS}(\sigma_4)={\cal C}_1\,{\cal A}_{\rm BAS}(1,2,3,4|\sigma_4)+{\cal C}_2\,{\cal A}_{\rm BAS}(1,3,2,4|\sigma_4)\,,~~~~\label{expan-nlsm-4p}
\eea
where the coefficients ${\cal C}_1$ and ${\cal C}_2$ have mass dimension $4$. Using \eref{4p1} and \eref{4p2}, we get equations
\bea
& &{\cal C}_1\,{\cal A}_{\rm BAS}(1,2,3,4|1,2,3,4)+{\cal C}_2\,{\cal A}_{\rm BAS}(1,3,2,4|1,2,3,4)=s+t\,,\nn
& &{\cal C}_1\,{\cal A}_{\rm BAS}(1,2,3,4|1,3,2,4)+{\cal C}_2\,{\cal A}_{\rm BAS}(1,3,2,4|1,3,2,4)=t+u\,,\nn
& &{\cal C}_1\,{\cal A}_{\rm BAS}(1,2,3,4|1,2,4,3)+{\cal C}_2\,{\cal A}_{\rm BAS}(1,3,2,4|1,2,4,3)=u+s\,.
\eea
After evaluating BAS amplitudes, the above equations are turned to
\bea
{\cal C}_1\,\Big({1\over s}+{1\over t}\Big)+{\cal C}_2\,\Big(-{1\over t}\Big)&=&s+t\,,\nn
{\cal C}_1\,\Big(-{1\over t}\Big)+{\cal C}_2\,\Big({1\over u}+{1\over t}\Big)&=&t+u\,,\nn
{\cal C}_1\,\Big(-{1\over s}\Big)+{\cal C}_2\,\Big(-{1\over u}\Big)&=&u+s\,.
\eea
It is straightforward to observe that only one equation of above three is independent, reflects the constraints from well known
KK and BCJ relations. Thus it is sufficient to solve the last one
\bea
{-{\cal C}_1\over s}+{-{\cal C}_2\over u}=u+s\,.~~~\label{eq-4pNLSM}
\eea

Based on the requirement that ${\cal C}_1$ and ${\cal C}_2$ are polynomials of Lorentz invariants without any pole (as discussed in subsection.\ref{subsecexpand} in section.\ref{sec-background}), the general solution to the equation \eref{eq-4pNLSM} is found to be
\bea
{\cal C}_1=\alpha\,s^2-(1+\beta)\,su\,,~~~~~~{\cal C}_2=\beta\,u^2-(1+\alpha)\,su\,.~~\label{solu-4pNLSM-general}
\eea
Different values of $\alpha$ and $\beta$ are related via BCJ relations, namely,
\bea
& &\Delta{\cal C}_1\,{\cal A}_{\rm BAS}(1,2,3,4|\sigma_4)+\Delta{\cal C}_2\,{\cal A}_{\rm BAS}(1,3,2,4|\sigma_4)\nn
&=&(s^2\,\Delta\alpha-su\,\Delta\beta)\,{\cal A}_{\rm BAS}(1,2,3,4|\sigma_4)+(u^2\,\Delta\beta-su\,\Delta\alpha)\,{\cal A}_{\rm BAS}(1,3,2,4|\sigma_4)\nn
&=&s\,\Delta\alpha\Big(s\,{\cal A}_{\rm BAS}(1,2,3,4|\sigma_4)-u\,{\cal A}_{\rm BAS}(1,3,2,4|\sigma_4)\Big)\nn
& &+u\,\Delta\beta\Big(u\,{\cal A}_{\rm BAS}(1,3,2,4|\sigma_4)-s\,{\cal A}_{\rm BAS}(1,2,3,4|\sigma_4)\Big)\nn
&=&0\,,~~\label{BCJ1}
\eea
where the last step uses the fundamental BCJ relation \eref{bcj}
\bea
0=s_{12}\,{\cal A}_{\rm BAS}(1,2,3,4|\sigma_4)+(s_{12}+s_{32})\,{\cal A}_{\rm BAS}(1,3,2,4|\sigma_4)\,.~~~\label{BCJ2}
\eea
It means different values of $\alpha$ and $\beta$ lead to equivalent physical amplitudes, thus one can fix this gauge by choosing special values of $\alpha$ and $\beta$. We require $\alpha=\beta$, to manifest the symmetry among $s$ and $u$ in equation \eref{eq-4pNLSM}. Then the general solution in \eref{solu-4pNLSM-general} is decomposed as the linear combination of two special ones which are
\bea
{\cal C}'_1=-s^2\,,~~~~~~{\cal C}'_2=-u^2\,,~~\label{solu1-4pNLSM}
\eea
and
\bea
{\cal C}''_1={\cal C}''_2=-su\,,~~\label{solu2-4pNLSM}
\eea
correspond to $\alpha=\beta=-1$ and $\alpha=\beta=0$, respectively. In the next subsection, we will fix the gauge further when extending the on-shell tree NLSM amplitudes to off-shell ones.

\subsection{Off-shell extension of tree NLSM amplitudes}
\label{subsec-offshell}

The lowest-point NLSM amplitudes have been determined in the previous subsection, and we want to construct higher-point amplitudes recursively from them, by inverting the universal soft theorem. To realize the goal, we need to find the desired soft behavior, which describe the factorization of the amplitude into the soft factor and the sub-amplitude with less external legs. However, since the $4$-point amplitudes are lowest-point ones, it is impossible to discuss such soft behavior without knowing any higher-point amplitudes. Then we encounter the difficulty that the method for constructing higher-point amplitudes requires knowing higher-point amplitudes first.

To solve this difficulty,
we extend the on-shell tree NLSM amplitudes to off-shell ones via the following procedure. First, we take external momenta $k_1$ and $k_n$ to be off-shell, i.e., $k_1^2\neq0$, $k_n^2\neq0$. Secondly, we assume that the off-shell NLSM amplitudes can also be expanded to off-shell KK BAS basis, where the off-shell BAS amplitudes are generated from on-shell ones by substituting off-shell momenta into the corresponding propagators.
In other words, the off-shell NLSM amplitudes are represented in the expanded formula
\bea
A_{\rm NLS}(\sigma_n)=\sum_{\sigma_{n-2}}\,C(\sigma_{n-2},k_i)\,A_{\rm BAS}(1,\sigma_{n-2},n|\sigma_n)\,.~~~\label{exp-N-KK-offshell}
\eea
This is the off-shell extension of the expansion in
\eref{exp-N-KK}, where the coefficients ${\cal C}(\sigma_{n-2},k_i)$ are replaced by the more general $C(\sigma_{n-2},k_i)$ which cover both on-shell and off-shell cases. We use $A$ to denote off-shell amplitudes, to distinguish them from the on-shell ones ${\cal A}$. Finally, we require the coefficients $C(\sigma_2,k_i)$ for $4$-point off-shell amplitudes should be reduced to those for the on-shell case when $k_1^2=k_4^2=0$. It is worth to emphasize that we do not require such off-shell amplitudes to be equivalent to those evaluated via Feynman rules with $k_1^2\neq0$, $k_n^2\neq0$. As will be seen, under such extension, it is natural to define the un-zero $3$-point NLSM amplitude, which vanishes in the special case $k_1^2=k_3^2=0$. Then we can find the desired soft behavior by decomposing the $4$-point amplitude into the soft factor and the $3$-point one.
In this subsection, we focus on the off-shell extension for the $3$-point and $4$-point amplitudes. The general off-shell amplitudes will be constructed in subsequent subsections by applying the recursive method.

As discussed in the previous subsection, the on-shell $3$-point tree NLSM amplitude does not exist due to kinematics. However, for the off-shell case with $k_1^2\neq0$, $k_3^2\neq0$, we have un-vanishing kinematical variables
\bea
k_1^2\,,~~k_3^2\,,~~k_1\cdot k_2={k_3^2-k_1^2\over 2}\,,~~k_1\cdot k_3=-{k_1^2+k_3^2\over2}\,,~~k_2\cdot k_3={k_1^2-k_3^2\over2}\,,~~\label{variable}
\eea
where $k_2$ is an on-shell massless momenta, satisfying the conservation condition $k_1+k_2+k_3=0$. These un-zero variables allow us to construct the un-vanishing off-shell extension of the $3$-point NLSM amplitude. To fix the explicit formula of such extended $3$-point amplitude, let us consider the expansion
\bea
A_{\rm NLS}(\sigma_3)=C\,A_{\rm BAS}(1,2,3|\sigma_3)\,,~~~\label{expan-3p}
\eea
with $k_1^2\neq0$, $k_3^2\neq0$. The natural extension of the basis ${\cal A}_{\rm BAS}(1,2,3|\sigma_3)$ is $A_{\rm BAS}(1,2,3|\sigma_3)={\cal A}_{\rm BAS}(1,2,3|\sigma_3)$ since it is independent of kinematics. More generally, we define $A_{\rm BAS}(\sigma'_3|\sigma_3)={\cal A}_{\rm BAS}(\sigma'_3|\sigma_3)$ for the arbitrary ordering $\sigma'_3$, then the relation ${\cal A}_{\rm BAS}(1,2,3|\sigma_3)=-{\cal A}_{\rm S}(3,2,1|\sigma_3)$ under the inversion of ordering still holds for the current off-shell case. This relation imposes $C'=-C$ when expanding $A_{\rm NLS}(\sigma_3)$ as
\bea
A_{\rm NLS}(\sigma_3)=C'\,A_{\rm BAS}(3,2,1|\sigma_3)\,.
\eea
Furthermore, the double copy structure indicates that $C$ depends on the ordering $(1,2,3)$ and $C'$ depends on $(3,2,1)$, both of them are independent of $\sigma_3$. Thus, $C'$ is generated from $C$ via the replacement $k_1\leftrightarrow k_3$, due to the relation among two orderings $(1,2,3)$ and $(3,2,1)$. Among Lorentz invariants in \eref{variable}, only $k_2\cdot k_1$ and $k_2\cdot k_3$
satisfy two conditions $C=-C'$ and $k_1\leftrightarrow k_3$ simultaneously. Thus, we find the appropriate choice
$C=\gamma\,k_2\cdot k_1$, $C'=\gamma\,k_2\cdot k_3$ where $\gamma$ is an arbitrary constant.

It is worth to emphasize that since $k_1^2\neq k_2^2$, $k_2^2\neq k_3^2$, and $k_1^2\neq k_3^2$, three external particles are no-longer un-distinguishable, thus the permutation symmetry among massless bosons is broken. For example, if one expand $A_{\rm NLS}(\sigma_3)$ as
\bea
A_{\rm NLS}(\sigma_3)=C''\,A_{\rm BAS}(1,3,2|\sigma_3)\,,
\eea
we do not expect that $C''$ can be generated from $C$ through the replacement $2\leftrightarrow3$.
However, the permutation $(1,2,3)\rightarrow(3,2,1)$ can be understood as re-defining the ordering by the opposite direction
without altering positions of external legs, thus we require that $C$ and $C'$ are related via the replacement $1\leftrightarrow 3$.

For later convenience, we choose $\gamma=2$ via the overall re-scaling of the amplitude, therefore
\bea
A_{\rm NLS}(\sigma_3)=(2\,k_2\cdot L_2)\,A_{\rm BAS}(1,2,3|\sigma_3)=(2\,k_2\cdot L_2)\,A_{\rm BAS}(3,2,1|\sigma_3)\,.~~~\label{expan-3p-final}
\eea
The combinatory momentum $L_i$ is defined as the summation of momenta carried by legs at the l.h.s of $i$ in the left ordering of the BAS amplitude. For $A_{\rm S}(1,2,3|\sigma_3)$, the left ordering is $(1,2,3)$ and the leg at the l.h.s of $2$ is $1$, thus we have $L_2=k_1$.
Similarly, for $A_{\rm S}(3,2,1|\sigma_3)$ we have $L_2=k_3$. One can define $R_i$ for the right ordering $\sigma_3$ in the analogous way.
Notice that $L$ and $R$ in above notations stand for left and right orderings respectively, rather than l.h.s and r.h.s of $i$.

Now we move to the extensions of $4$-point amplitudes. In the expanded formula \eref{expan-nlsm-4p}, the BAS amplitudes
${\cal A}_{\rm BAS}(1,2,3,4|\sigma_4)$ and ${\cal A}_{\rm BAS}(1,3,2,4|\sigma_4)$ can be extended directly by replacing Mandelstam variables in propagators by those with off-shell $k_1$ and $k_4$, namely,
\bea
s&=&k_1^2+2\,k_1\cdot k_2=k_4^2+2\,k_3\cdot k_4\,,\nn
t&=&k_1^2+k_4^2+2\,k_1\cdot k_4=2\,k_2\cdot k_3\,,\nn
u&=&k_1^2+2\,k_1\cdot k_3=k_4^2+2\,k_2\cdot k_4\,.
\eea
The challenge is the extension of coefficients ${\cal C}_1$ and ${\cal C}_2$ to off-shell ones $C_1$ and $C_2$.
The difficulty arises from the following ambiguity. For the on-shell case, the Mandelstam variables have various equivalent expressions,
\bea
s&=&(k_1+k_2)^2=(k_3+k_4)^2=2\,k_1\cdot k_2=2\,k_3\cdot k_4\,,\nn
t&=&(k_1+k_4)^2=(k_2+k_3)^2=2\,k_1\cdot k_4=2\,k_2\cdot k_3\,,\nn
u&=&(k_1+k_3)^2=(k_2+k_4)^2=2\,k_1\cdot k_3=2\,k_2\cdot k_4\,.~~\label{choices}
\eea
For the off-shell case with $k_1^2\neq0$, $k_4^2\neq0$, the equivalences
\bea
s&=&(k_1+k_2)^2=(k_3+k_4)^2\,,\nn
t&=&(k_1+k_4)^2=(k_2+k_3)^2\,,\nn
u&=&(k_1+k_3)^2=(k_2+k_4)^2\,
\eea
still hold, while others are spoiled. We need to choose appropriate formulas among these inequivalent candidates to define $C_1$ and $C_2$, since all these candidates are reduced to the correct ${\cal C}_1$ and ${\cal C}_2$ when $k_1$ and $k_4$ are on-shell. Furthermore, if we modify $C_1$ and $C_2$ as
\bea
C'_1&=&C_1+\alpha_1\,(k_1^2)^2+\beta_1\,(k_4^2)^2+\gamma_1\,k_1^2\,k_4^2\,\nn
C'_2&=&C_2+\alpha_2\,(k_1^2)^2+\beta_2\,(k_4^2)^2+\gamma_2\,k_1^2\,k_4^2\,,~~~\label{modi-abr}
\eea
the resulted $C'_1$ and $C'_2$ are inequivalent to $C_1$ and $C_2$, but are also reduced to ${\cal C}_1$ and ${\cal C}_2$ when taking the on-shell limit $k_1^2=k_4^2=0$.

The unique choice for $C_1$ and $C_2$ can be made by considering poles. Although $A_{\rm NLS}(\sigma_4)$ can not be interpreted as the physical scattering amplitudes, we hope that they inherit some general features of amplitudes which allow gluing lower-point objects together to get higher-point ones, otherwise the bottom up recursive construction will be inconsistent. This expectation implies that the extended off-shell objects should have appropriate structure of poles. Suppose $A_{\rm NLS}(\sigma_4)$ contains a propagator, then it factorizes into two $3$-point amplitudes $A_L$ and $A_R$ at the corresponding pole. Since the on-shell $3$-point amplitudes do not exist, two off-shell legs must be separated in the factorization, i.e., one of them belongs to $A_L$ and another one belongs to $A_R$. Consequently, $A_{\rm NLS}(1,2,3,4)$ has a propagator $1/s_{12}$, $A_{\rm NLS}(1,3,2,4)$ has a propagator $1/s_{13}$, while $A_{\rm NLS}(1,2,4,3)$ includes two propagators $1/s_{12}$ and $1/s_{13}$. We can also consider a special case that $k_1$ or $k_4$ is on-shell, leaving only one momentum to be off-shell. For such special case, $A_{\rm NLS}(\sigma_4)$ has only one off-shell external leg, one of two $3$-point amplitudes in factorization must be on-shell thus vanishing due to kinematics. Thus, if only one of $k_1$ and $k_4$ is off-shell, all $A_{\rm NLS}(\sigma_4)$ are polynomials of Mandelstam variables without any pole. The coefficients $C_1$ and $C_2$ will be uniquely determined by imposing above requirements for poles.

As discussed in the previous subsection, the general solution of ${\cal C}_1$ and ${\cal C}_2$
in \eref{expan-nlsm-4p} can be decomposed as the combination of two special solutions in \eref{solu1-4pNLSM} and \eref{solu2-4pNLSM}. We focus on the solution \eref{solu1-4pNLSM} first. For the on-shell case, ${\cal C}'_1$ and ${\cal C}'_2$ have following equivalent formulas,
\bea
& &{\cal C}'_1=-s^2=-(2\,k_1\cdot k_2)^2=-(2\,k_3\cdot k_4)^2=-(2\,k_1\cdot k_2)\,(2\,k_3\cdot k_4)\,,\nn
& &{\cal C}'_2=-u^2=-(2\,k_1\cdot k_3)^2=-(2\,k_2\cdot k_4)^2=-(2\,k_1\cdot k_3)\,(2\,k_2\cdot k_4)\,,~~\label{choice-onshell}
\eea
as can be seen by substituting \eref{choices} into \eref{solu1-4pNLSM}.
Among these candidates, the appropriate choice for the off-shell case is
\bea
& &C'_1=-(2\,k_1\cdot k_2)\,(2\,k_3\cdot k_4)\,,\nn
& &C'_2=-(2\,k_1\cdot k_3)\,(2\,k_2\cdot k_4)\,.~~~\label{choice-1}
\eea
When $k_1^2\neq0$, $k_4^2\neq0$, this choice leads to
\bea
& &A_{\rm NLS}(1,2,3,4)=2\,k_1\cdot (k_2-k_3)-{(2\,k_1\cdot k_2)\,(2\,k_3\cdot k_4)\over s}\,,\nn
& &A_{\rm NLS}(1,3,2,4)=2\,k_1\cdot (k_3-k_2)-{(2\,k_1\cdot k_3)\,(2\,k_2\cdot k_4)\over u}\,,\nn
& &A_{\rm NLS}(1,2,4,3)={(2\,k_1\cdot k_2)\,(2\,k_3\cdot k_4)\over s}+{(2\,k_1\cdot k_3)\,(2\,k_2\cdot k_4)\over u}\,,~~\label{4p-1offshell}
\eea
satisfying the expectation for poles.
When one of two momenta $k_1$ and $k_4$ is on-shell, poles in above results are canceled. For example, imposing $k_4^2=0$ yields
\bea
& &A_{\rm NLS}(1,2,3,4)=-2\,k_3\cdot k_1\,,\nn
& &A_{\rm NLS}(1,3,2,4)=-2\,k_2\cdot k_1\,,\nn
& &A_{\rm NLS}(1,2,4,3)=2\,k_1\cdot k_{23}\,,~~\label{4p-1offshell}
\eea
all $A_{\rm NLS}(\sigma_4)$ are polynomials without any pole, as we expected.
One can verify that other choices in \eref{choice-onshell} lead to incorrect structure of poles. For example, in the special case $k_4^2=0$, the choice $C'_1=-s^2$ and $C'_2=-u^2$ yields
\bea
& &A_{\rm NLS}(1,2,3,4)=2{(k_4\cdot(k_3-k_2))\,(k_4\cdot k_1)\over k_2\cdot k_3}-s\,,\nn
& &A_{\rm NLS}(1,3,2,4)=2{(k_4\cdot(k_2-k_3))\,(k_4\cdot k_1)\over k_2\cdot k_3}-u\,,\nn
& &A_{\rm NLS}(1,2,4,3)=s+u\,,
\eea
$A_{\rm NLS}(1,2,3,4)$ and $A_{\rm NLS}(1,3,2,4)$ contain poles thus are un-acceptable. Similar argument also excludes the modifications in \eref{modi-abr}.

Then we consider the solution \eref{solu2-4pNLSM}.
When extending to the off-shell case, both $s$ and $u$ have three inequivalent choices in \eref{choices}. However, the appropriate choice satisfying the requirements for poles does not exist. The choice
\bea
C''_1=C''_2=-su~~~~\label{choice-2}
\eea
leads to
\bea
& &A_{\rm NLS}(1,2,3,4)=-s\,,\nn
& &A_{\rm NLS}(1,3,2,4)=-u\,,\nn
& &A_{\rm NLS}(1,2,4,3)=u+s\,,
\eea
which do not contain any pole when $k_1^2\neq0$, $k_4^2\neq0$. For other choices, the resulting $A_{\rm NLS}(\sigma_4)$ contain poles when $k_1^2=0$
or $k_4^2=0$.

Consequently, the coefficients $C_1$ and $C_2$ which satisfy all requirements for poles are uniquely fixed as in \eref{choice-1}. Thus, the off-shell extension of ${\cal A}_{\rm NLS}(\sigma_4)$ is
\bea
A_{\rm NLS}(\sigma_4)&=&-4\,(k_1\cdot k_2)\,(k_3\cdot k_4)\,A_{\rm BAS}(1,2,3,4|\sigma_4)-4\,(k_1\cdot k_3)\,(k_2\cdot k_4)\,A_{\rm BAS}(1,3,2,4|\sigma_4)\nn
&=&(2\,k_2\cdot L_2)\,(2\,k_3\cdot L_3)\,A_{\rm BAS}(1,2\shuffle3,4|\sigma_4)\,,~~~\label{expan-nlsm-4p-offshell}
\eea
where the meaning of $\shuffle$ was explained in subsection.\ref{subsecexpand} in section.\ref{sec-background}, and the combinatory momenta $L_i$
were defined after \eref{expan-3p-final}.

\subsection{Recursive construction and $5$-point amplitude}
\label{subsec-5p}

In this subsection, we construct the $5$-point off-shell NLSM amplitudes from the $4$-point ones.
The recursive method based on the assumption of the universality of single soft behavior, thus the first step is to figure out the soft factor.
Let us re-scale $k_2$ as $k_2\to\tau k_2$, and expand $A_{\rm NLS}(\sigma_4)$ in \eref{expan-nlsm-4p-offshell} by $\tau$. The leading order contribution is found to be
\bea
A^{(0)_2}_{\rm NLS}(\sigma_4)&=&(2\,k_2\cdot k_1)\,(2\,k_3\cdot k_1)\,{\delta_{23}\over s_{23}}\,A_{\rm BAS}(1,{\not 2},3,4|\sigma_4\setminus2)\nn
& &+(2\,k_2\cdot k_{13})\,(2\,k_3\cdot k_1)\,{\delta_{32}\over s_{32}}\,A_{\rm BAS}(1,3,{\not 2},4|\sigma_4\setminus2)\nn
&=&\delta_{32}\,(2\,k_3\cdot k_1)\,A_{\rm BAS}(1,3,4|\sigma_4\setminus2)\nn
&=&\delta_{32}\,A_{\rm NLS}(\sigma_4\setminus2)\,,~~\label{soft1-4p}
\eea
where the last equality uses the off-shell $3$-point amplitude in \eref{expan-3p-final}. The soft behavior of off-shell tree BAS amplitudes
are obtained by eliminating $\delta_{12}/s_{12}$ and $\delta_{24}/s_{24}$ in \eref{soft-theo-s} and \eref{soft-fac-s-0}, as discussed in the end of subsection.\ref{subsecBAS} in section.\ref{sec-background}.
Similarly, one can re-scale $k_3$ and find
\bea
A^{(0)_3}_{\rm NLS}(\sigma_4)
&=&\delta_{23}\,A_{\rm NLS}(\sigma_4\setminus3)\,,~~~\label{soft2-4p}
\eea
From \eref{soft1-4p} and \eref{soft2-4p}, we observe the universal soft behavior
\bea
A^{(0)_i}_{\rm NLS}(\sigma_4)
&=&S^{(0)_i}_{\rm NLS}\,A_{\rm NLS}(\sigma_4\setminus i)\,,~~~\label{soft-4p}
\eea
where
\bea
S^{(0)_i}_{\rm NLS}=\sum_{j\neq1,4,i}\,\delta_{ji}\,.~~\label{soft-fac-N-4p}
\eea
Then the universality of soft factor indicates that
\bea
A^{(0)_i}_{\rm NLS}(\sigma_n)
&=&S^{(0)_i}_{\rm NLS}\,A_{\rm NLS}(\sigma_n\setminus i)\,,~~~\label{soft-np}
\eea
where
\bea
S^{(0)_i}_{\rm NLS}=\sum_{j\neq1,n,i}\,\delta_{ji}\,,~~\label{soft-fac-N-0}
\eea
for general $A_{\rm NLS}(\sigma_n)$ with $n$ external legs.
The reason for introducing the summation over $j$ in \eref{soft-fac-N-4p} and \eref{soft-fac-N-0} is the symmetry:
all legs $j\in\{2,\cdots,n-1\}$ are on-shell scalars, there is no reason to select a particular subset of them and neglect others.

By imposing the universal soft behavior in \eref{soft-np} and \eref{soft-fac-N-0}, we can construct the $5$-point off-shell amplitudes
$A_{\rm NLS}(\sigma_5)$ from $4$-point ones in \eref{expan-nlsm-4p-offshell}. Re-scaling the external momentum $k_2\to\tau k_2$ and expand $A_{\rm NLS}(\sigma_5)$ by $\tau$, the soft theorem in \eref{soft-np} and \eref{soft-fac-N-0} requires
\bea
A_{\rm NLS}^{(0)_2}(\sigma_5)&=&\Big(\delta_{32}+\delta_{42}\Big)\,A_{\rm NLS}(\sigma_5\setminus 2)\nn
&=&\Big(\delta_{32}+\delta_{42}\Big)\,\Big[(2\,k_3\cdot L_3)\,(2\,k_4\cdot L_4)\,A_{\rm BAS}(1,3\shuffle4,5|\sigma_5\setminus2)\Big]\nn
&=&(2\,k_3\cdot L_3)\,(2\,k_4\cdot L_4)\,\Big((2\,k_2\cdot k_3)\,{\delta_{32}\over s_{23}}+(2\,k_2\cdot k_4){\delta_{42}\over s_{24}}\Big)\,A_{\rm BAS}(1,3\shuffle4,5|\sigma_5\setminus2)\nn
&=&P_1+P_2\,,~~\label{p1p2}
\eea
where
\bea
P_1&=&(2\,k_3\cdot L_3)\,(2\,k_4\cdot L_4)\,\Big[(2\,k_2\cdot X_{(3,4)_2})\,{\delta_{23}\over s_{23}}+(2\,k_2\cdot(X_{(3,4)_2}+k_3))\,{\delta_{32}\over s_{23}}\nn
& &+(2\,k_2\cdot(X_{(3,4)_2}+k_3))\,{\delta_{24}\over s_{24}}+(2\,k_2\cdot(X_{(3,4)_2}+k_{34}))\,{\delta_{42}\over s_{24}}\Big]\,A_{\rm BAS}(1,3,4,5|\sigma_5\setminus2)\,,~~\label{P1}
\eea
and
\bea
P_2& =&(2\,k_3\cdot L_3)\,(2\,k_4\cdot L_4)\,\Big[(2\,k_2\cdot X_{(4,3)_2})\,{\delta_{24}\over s_{24}}+(2\,k_2\cdot(X_{(4,3)_2}+k_4))\,{\delta_{42}\over s_{24}}\nn
& &+(2\,k_2\cdot(X_{(4,3)_2}+k_4))\,{\delta_{23}\over s_{23}}+(2\,k_2\cdot(X_{(4,3)_2}+k_{43}))\,{\delta_{32}\over s_{32}}\Big]\,A_{\rm BAS}(1,4,3,5|\sigma_5\setminus2)\,.~~\label{P2}
\eea
Here $X_{(3,4)_2}$ is an un-fixed momentum without including $k_2$, since the component proportional to $k_2$ does not contribute to $k_2\cdot X_{(3,4)_2}$. The subscript $(3,4)_2$ means $X_{(3,4)_2}$ is associated with the ordering $(1,3,4,5)$ when considering the soft behavior of external leg $2$. The un-fixed momentum $X_{(4,3)_2}$ is analogous. Substituting the leading order soft behavior of off-shell BAS amplitudes, we find
\bea
P_1&=&(2\,k_3\cdot L^{\not 2}_3)\,(2\,k_4\cdot L^{\not 2}_4)\,\Big[(2\,k_2\cdot X_{(3,4)_2})\,A^{(0)_2}_{\rm BAS}(1,2,3,4,5|\sigma_5)\nn
& &+(2\,k_2\cdot(X_{(3,4)_2}+k_3))\,A^{(0)_2}_{\rm BAS}(1,3,2,4,5|\sigma_5)\nn
& &+(2\,k_2\cdot(X_{(3,4)_2}+k_{34}))\,A^{(0)_2}_{\rm BAS}(1,3,4,2,5|\sigma_5)\Big]\,,~~\label{p1}
\eea
and
\bea
P_2&=&(2\,k_3\cdot L^{\not 2}_3)\,(2\,k_4\cdot L^{\not 2}_4)\,\Big[(2\,k_2\cdot X_{(4,3)_2})\,A^{(0)_2}_{\rm BAS}(1,2,4,3,5|\sigma_5)\nn
& &+(2\,k_2\cdot(X_{(4,3)_2}+k_4))\,A^{(0)_2}_{\rm BAS}(1,4,2,3,5|\sigma_5)\nn
& &+(2\,k_2\cdot(X_{(4,3)_2}+k_{43}))\,A^{(0)_2}_{\rm BAS}(1,4,3,2,5|\sigma_5)\Big]\,.~~\label{p2}
\eea
We emphasize that $L_3$ and $L_4$ in \eref{P1} and \eref{P2} are defined for orderings $(1,3,4,5)$ and $(1,4,3,5)$ with four legs,
while in \eref{p1} and \eref{p2} are defined for ordering such as $(1,2,3,4,5)$ with five legs. The superscript $\not 2$ means delating
$k_2$ in $L_3$ and $L_4$. Substituting \eref{p1} and \eref{p2} into \eref{p1p2}, we arrive at
\bea
A^{(0)_2}_{\rm NLS}(\sigma_5)&=&(2\,k_3\cdot L^{\not 2}_3)\,(2\,k_4\cdot L^{\not 2}_4)\,(2\,k_2\cdot (L_2-k_1+X_{(3,4)_2}))\,A^{(0)_2}_{\rm BAS}(1,2\shuffle\{3,4\},5|\sigma_5)\nn
& &+(2\,k_3\cdot L^{\not 2}_3)\,(2\,k_4\cdot L^{\not 2}_4)\,(2\,k_2\cdot (L_2-k_1+X_{(4,3)_2}))\,A^{(0)_2}_{\rm BAS}(1,2\shuffle\{4,3\},5|\sigma_5)\,.~~\label{5p-2soft}
\eea
Similar manipulations hold for $k_3\to\tau k_3$ and $k_4\to\tau k_4$, which lead to
\bea
A^{(0)_3}_{\rm NLS}(\sigma_5)&=&(2\,k_2\cdot L^{\not 3}_2)\,(2\,k_4\cdot L^{\not 3}_4)\,(2\,k_3\cdot (L_3-k_1+X_{(2,4)_3}))\,A^{(0)_3}_{\rm BAS}(1,3\shuffle\{2,4\},5|\sigma_5)\nn
& &+(2\,k_2\cdot L^{\not 3}_2)\,(2\,k_4\cdot L^{\not 3}_4)\,(2\,k_3\cdot (L_3-k_1+X_{(4,2)_3}))\,A^{(0)_3}_{\rm BAS}(1,3\shuffle\{4,2\},5|\sigma_5)\,,~~\label{5p-3soft}
\eea
as well as
\bea
A^{(0)_4}_{\rm NLS}(\sigma_5)&=&(2\,k_2\cdot L^{\not 4}_2)\,(2\,k_3\cdot L^{\not 4}_3)\,(2\,k_4\cdot (L_4-k_1+X_{(2,3)_4}))\,A^{(0)_4}_{\rm BAS}(1,4\shuffle\{2,3\},5|\sigma_5)\nn
& &+(2\,k_2\cdot L^{\not 4}_2)\,(2\,k_3\cdot L^{\not 4}_3)\,(2\,k_4\cdot (L_4-k_1+X_{(3,2)_4}))\,A^{(0)_4}_{\rm BAS}(1,4\shuffle\{3,2\},5|\sigma_5)\,.~~\label{5p-4soft}
\eea

The leading order soft behavior can also be expressed as
\bea
A^{(0)_i}_{\rm NLS}(\sigma_5)=\sum_{\sigma_3}\,C^{(0)_i}(\sigma_3)\,A^{(0)_i}_{\rm BAS}(1,\sigma_3,5|\sigma_5)\,,~~\label{soft-5p}
\eea
due to the basic ansatz in \eref{exp-N-KK-offshell}. Here $\sigma_3$ denotes the orderings among external legs in $\{2,3,4\}$.
The coefficients $C^{(0)_i}(\sigma_3)$ are generated from $C(\sigma_3)$ as follows. The mass dimension
indicates that $C(\sigma_3)$ can be decomposed as
\bea
C(\sigma_3)=({\cal K}_1\cdot{\cal K}_4)\,({\cal K}_2\cdot{\cal K}_5)\,({\cal K}_3\cdot{\cal K}_6)\,,
\eea
where each ${\cal K}_j$ is the combination of external momenta and depends on the ordering $\sigma_3$.
Then we have
\bea
C^{(0)_i}(\sigma_3)=({\cal K}^{(0)_i}_1\cdot{\cal K}^{(0)_i}_2)\,({\cal K}^{(0)_i}_3\cdot{\cal K}^{(0)_i}_4)\,({\cal K}^{(0)_i}_5\cdot{\cal K}^{(0)_i}_6)\,.
\eea
Obviously, ${\cal K}^{(0)_i}_j$ are generated from ${\cal K}_j$ via the rules
\bea
{\cal K}^{(0)_i}_j=\begin{cases}\displaystyle ~\tau k_i~~~~ &{\rm if}~{\cal K}_j=k_i \,,\\
\displaystyle ~{\cal K}^{\not i}_j~~~~~~~ & {\rm otherwise}\,.\end{cases}~~~\label{defin-calk}
\eea
Again, the superscript $\not i$ means delating the component proportional to $k_i$. Comparing \eref{soft-5p}
with \eref{5p-2soft}, we see that $C(\sigma_3)$ contains a factor $(2\,k_2\cdot (L_2-k_1+X_{(\sigma_3\setminus2)_2}))$,
due to the rules in \eref{defin-calk}.
Similarly, comparing \eref{soft-5p} with \eref{5p-3soft} and \eref{5p-4soft} indicates factors $(2\,k_3\cdot (L_3-k_1+X_{(\sigma_3\setminus3)_3}))$
and $(2\,k_4\cdot (L_4-k_1+X_{(\sigma_3\setminus4)_4}))$. Three factors provide the correct mass dimension of $C(\sigma_3)$, therefore
\bea
C(\sigma_3)=\prod_{i=2,3,4}\,2\,k_i\cdot (L_i-k_1+X_{(\sigma_3\setminus i)_i})\,.~~\label{c-5p-1}
\eea

To fix the momenta $X_{(\sigma_3\setminus i)_i}$, one can use $C(\sigma_3)$ in \eref{c-5p-1} to evaluate $C^{(0)_i}(\sigma_3)$ and compare the results with \eref{5p-2soft}, \eref{5p-3soft} and \eref{5p-4soft}. Such manipulation gives $X_{(\sigma_3\setminus i)_i}=k_1$, for all $\sigma_3$ and $i$.
For instance, \eref{c-5p-1} leads to
\bea
C^{(0)_2}(2,3,4)=\Big(2\,k_2\cdot (L_2-k_1+X_{(3,4)_2})\Big)\,\Big(2\,k_3\cdot (L^{(\not2}_3-k_1+X^{(0)_ 2}_{(2,4)_3})\Big)\,\Big(2\,k_4\cdot (L^{\not 2}_4-k_1+X^{(0)_2}_{(2,3)_4})\Big)\,.~~\label{step1}
\eea
Comparing it with \eref{5p-2soft}, we find the constraints
\bea
X^{(0)_2}_{(2,4)_3}=k_1\,,~~~~X^{(0)_2}_{(2,3)_4}=k_1\,,
\eea
therefore
\bea
X_{(2,4)_3}-k_1=a_{(2,4)_3}\, k_2\,,~~~~~X_{(2,3)_4}-k_1=a_{(2,3)_4}\,k_2\,,~~\label{eq-1}
\eea
where $a_{(2,4)_3}$ and $a_{(2,3)_4}$ are two constants. On the other hand, \eref{c-5p-1} also gives
\bea
C^{(0)_3}(2,3,4)=\Big(2\,k_3\cdot (L_3-k_1+X_{(2,4)_3})\Big)\,\Big(2\,k_2\cdot (L^{\not 3}_2-k_1+X^{(0)_3}_{(3,4)_2})\Big)\,\Big(2\,k_4\cdot (L^{\not 3}_4-k_1+X^{(0)_3}_{(2,3)_4})\Big)\,,
\eea
and comparing it with \eref{5p-3soft} yields
\bea
X_{(3,4)_2}-k_1=b_{(3,4)_2}\,k_3\,,~~~~X_{(2,3)_4}-k_1=b_{(2,3)_4}\,k_3\,.~~\label{eq-2}
\eea
Combining equations in \eref{eq-1} and \eref{eq-2} together, we get
\bea
a_{(2,3)_4}\,k_2=b_{(2,3)_4}\,k_3\,,~~\label{stepf}
\eea
and the solution is $a_{(2,3)_4}=b_{(2,3)_4}=0$ due to the independence of $k_2$ and $k_3$. This solution
indicates $X_{(2,3)_4}=k_1$. Other $X_{(\sigma_3\setminus i)_i}$ can be determined via the similar argument.
Consequently, the coefficients $C(\sigma_3)$ are completely fixed as
\bea
C(\sigma_3)=\prod_{i=2,3,4}\,2\,k_i\cdot L_i\,,~~\label{c-5p-2}
\eea
and the $5$-point off-shell amplitudes $A_{\rm N}(\sigma_5)$ are determined as
\bea
A_{\rm NLS}(\sigma_5)=\sum_{\sigma_3}\,\Big(\prod_{i=2,3,4}\,2\,k_i\cdot L_i\Big)\,A_{\rm BAS}(1,\sigma_3,5|\sigma_5)\,.~~\label{expan-5p-final}
\eea
%

\subsection{$n$-point amplitudes and enhanced Adler zero}
\label{subsec-np}

In the previous subsection, we have generated the $5$-point amplitudes from the $4$-point ones. Such recursive method can be applied to the general
case straightforwardly. Let us assume that the $m$-point off-shell tree NLSM amplitudes can be expanded as
\bea
A_{\rm NLS}(\sigma_m)=\sum_{\sigma_{m-2}}\,\Big(\prod_{i=2}^{m-1}\,2\,k_i\cdot L_i\Big)\,A_{\rm BAS}(1,\sigma_{m-2},m|\sigma_m)\,.~~\label{expan-mp}
\eea
Obviously, all $3$-point, $4$-point and $5$-point amplitudes in \eref{expan-3p-final}, \eref{expan-nlsm-4p-offshell} and \eref{expan-5p-final} satisfy this formula. Our purpose is to construct the $(m+1)$-point amplitudes $A_{\rm NLS}(\sigma_{m+1})$ which carry external legs in $\{1,\cdots,m+1\}$. The process is parallel to that for constructing $A_{\rm NLS}(\sigma_5)$ in the previous subsection. We re-scale the external momenta $k_i$ as $k_i\to\tau k_i$, where $i\in\{2,\cdots,m-1\}$, and expand $A_{\rm NLS}(\sigma_{m+1})$ by $\tau$. The universal soft behavior in \eref{soft-np} and \eref{soft-fac-N-0} requires
\bea
A_{\rm NLS}^{(0)_i}(\sigma_{m+1})&=&\Big(\sum_{j\neq 1,m+1,i}\,\delta_{ji}\Big)\,A_{\rm NLS}(\sigma_{m+1}\setminus i)\nn
&=&\Big(\sum_{j\neq 1,m+1,i}\,\delta_{ji}\Big)\,\Big[\sum_{\sigma_{m-2}}\,\Big(\prod_{\ell\in\{2,\cdots,m\}\setminus i}\,2\,k_\ell\cdot L_\ell\Big)\,A_{\rm BAS}(1,\sigma_{m-2},m+1|\sigma_{m+1}\setminus i)\Big]\nn
&=&\sum_{\sigma_{m-2}}\,\Big(\prod_{\ell\in\{2,\cdots,m\}\setminus i}\,2\,k_\ell\cdot L_\ell\Big)\,\Big(\sum_{j\neq 1,m+1,i}\,(2\,k_j\cdot k_i)\,{\delta_{ji}\over s_{ji}}\Big)\,A_{\rm BAS}(1,\sigma_{m-2},m+1|\sigma_{m+1}\setminus i)\nn
&=&\sum_{\sigma_{m-1}}\,\Big(\prod_{\ell\in\{2,\cdots,m\}\setminus i}\,2\,k_\ell\cdot L^{\not i}_\ell\Big)\,\Big(2\,k_i\cdot(L_i-k_1+X_{(\sigma_{m-1}\setminus i)_i})\Big)\,A^{(0)_i}_{\rm BAS}(1,\sigma_{m-1},m+1|\sigma_{m+1})\,,~~\label{soft-m+1}
\eea
where $X_{(\sigma_{m-1}\setminus i)_i}$ denotes the un-fixed momentum without including $k_i$.
Here orderings $\sigma_{m-2}$ in second and third lines are among legs in $\{2,\cdots,m\}\setminus i$, while $\sigma_{m-1}$ in the last line associates with legs in $\{2,\cdots,m\}$. The combinatory momenta $L_\ell$ in second and third lines are defined for the ordering $(1,\sigma_{m-2},m+1)$,
while in the last line are defined for $(1,\sigma_{m-1},m+1)$. The calculation for the last line is parallel to those in \eref{P1}, \eref{P2}, \eref{p1}, \eref{p2} and \eref{5p-2soft}. The ansatz in \eref{exp-N-KK-offshell} indicates that the leading order contribution $A_{\rm NLS}^{(0)_i}(\sigma_{m+1})$ can be expressed as
\bea
A^{(0)_i}_{\rm NLS}(\sigma_{m+1})=\sum_{\sigma_{m-1}}\,C^{(0)_i}(\sigma_{m-1})\,A^{(0)_i}_{\rm BAS}(1,\sigma'_{m-1},m+1|\sigma_{m+1})\,,~~\label{A0}
\eea
and the correct mass dimension indicates the decomposition
\bea
C(\sigma_{m-1})&=&\prod_{j=1}^{m-1}\,{\cal K}_j\cdot {\cal K}_{j+m-1}\,,\nn
C^{(0)_i}(\sigma_{m-1})&=&\prod_{j=1}^{m-1}\,{\cal K}^{(0)_i}_j\cdot {\cal K}^{(0)_i}_{j+m-1}\,,~~~\label{c0}
\eea
where the combinatory momenta ${\cal K}^{(0)_i}_a$ are generated from ${\cal K}_a$ via the rules in \eref{defin-calk}.
Substituting \eref{c0} into \eref{A0} and comparing with \eref{soft-m+1}, we observe
\bea
C(\sigma_{m-1})=\prod_{i=2}^m\,2\,k_i\cdot(L_i-k_1+X_{(\sigma_{m-1}\setminus i)_i})\,,
\eea
satisfying
\bea
L^{(0)_j}_i-k_1+X^{(0)_j}_{(\sigma_{m-1}\setminus i)_i}=L^{(0)_j}_i\,,~~~~~~{\rm for}~j\neq i\,.~~\label{equ-NLSM}
\eea
Using equation \eref{equ-NLSM}, one can fix the momenta $X_{(\sigma_{m-1}\setminus i)_i}$ as $X_{(\sigma_{m-1}\setminus i)_i}=k_1$,
the argument is the same as that from \eref{step1} to \eref{stepf} in the previous subsection. Thus the $(m+1)$-point amplitudes are uniquely determined in the expanded formula
\bea
A_{\rm NLS}(\sigma_{m+1})=\sum_{\sigma_{m-1}}\,\Big(\prod_{i=2}^m\,2\,k_i\cdot L_i\Big)\,A_{\rm BAS}(1,\sigma_{m-1},m+1|\sigma_{m+1})\,.~~\label{expan-m+1p}
\eea
From expressions in \eref{expan-mp} and \eref{expan-m+1p}, we conclude that the general off-shell tree NLSM amplitude
with $n$ external legs can be expanded as
\bea
A_{\rm NLS}(\sigma_n)=\sum_{\sigma_{n-2}}\,\Big(\prod_{i=2}^{n-1}\,2\,k_i\cdot L_i\Big)\,A_{\rm BAS}(1,\sigma_{n-2},n|\sigma_n)\,.~~\label{expan-offshellNLSM-np}
\eea

The physical on-shell amplitudes can be obtained from \eref{expan-offshellNLSM-np} by setting $k_1^2=k_n^2=0$. In the on-shell limit, expansion in \eref{expan-offshellNLSM-np} with odd number of external particles vanish, due to the BCJ relations, as can be seen in \cite{Carrasco:2016ldy,Du:2018khm}. For on-shell amplitudes with even number of external legs, the expanded formula is inherited from \eref{expan-offshellNLSM-np}, i.e.,
\bea
{\cal A}_{\rm NLS}(\sigma_n)=\sum_{\sigma_{n-2}}\,\Big(\prod_{i=2}^{n-1}\,2\,k_i\cdot L_i\Big)\,{\cal A}_{\rm BAS}(1,\sigma_{n-2},n|\sigma_n)\,,~~\label{expan-onshellNLSM-np}
\eea
coincide with the result found in \cite{Feng:2019tvb} by using differential operators.

For the on-shell physical NLSM amplitudes in the expanded formula in \eref{expan-onshellNLSM-np}, the most interesting observation is the enhanced Adler zero, i.e., the soft behavior of the on-shell amplitude ${\cal A}_{\rm NLS}(\sigma_n)$ vanish at the $\tau^0$ order. From the top down perspective based on Lagrangian, the enhanced Adler zero means the amplitudes have vanished soft behavior which exceed the degree expected from naive derivative power counting. For the current NLSM case, the number of derivatives per field in Lagrangian implies that the amplitudes vanish at the   $\tau^{-1}$ order, but actually they also vanish at the $\tau^0$ order. In this paper, we use the bottom up method to construct the general NLSM amplitudes with out the aid of a Lagrangian. Thus, we should describe enhanced Alder zero in the following slightly different way. From the expansion in \eref{expan-onshellNLSM-np}, we see that the leading order of coefficients $\prod_{i=2}^{n-1}\,2\,k_i\cdot L_i$ is the $\tau^1$ order, for any external momentum $k_i$ which is taken to be soft. At the same time, the leading order of BAS basis is the $\tau^{-1}$ order. Thus, the naive power counting for the expanded amplitudes in \eref{expan-onshellNLSM-np} leads to the expectation that ${\cal A}^{(0)_i}_{\rm NLS}(\sigma_n)$ is at the $\tau^{0}$ order. However, using the expansion \eref{expan-onshellNLSM-np}, it is very straightforward to observe the vanishing of ${\cal A}^{[\tau^0]_i}_{\rm NLS}(\sigma_n)$, where the superscript $[\tau^0]_i$ denotes the $\tau^0$ order contribution when $k_i\to\tau k_i$.
Substituting the soft theorem for BAS amplitudes in \eref{soft-theo-s} and \eref{soft-fac-s-0} into \eref{expan-onshellNLSM-np}, one can calculate the soft behavior of on-shell ${\cal A}_{\rm NLS}(\sigma_n)$, the process is parallel to that in \eref{soft1-4p}. The resulted soft behavior is
\bea
{\cal A}^{[\tau^0]_i}_{\rm NLS}(\sigma_n)=\Big(\sum_{j\neq i}\,\delta_{ji}\Big)\,\Big[\lim_{k_1^2\to 0,k_n^2\to 0}\,A_{\rm NLS}(\sigma_n\setminus i)\Big]\,.~~\label{nlsm-t0}
\eea
In \eref{nlsm-t0}, two factors $\sum_{j\neq i}\,\delta_{ji}$ and $\lim_{k_1^2\to 0,k_n^2\to 0}\,A_{\rm NLS}(\sigma_n\setminus i)$ vanish simultaneously due to the identity \eref{iden} and the odd number of external legs respectively. Thus we find that ${\cal A}^{[\tau^0]_i}_{\rm NLS}(\sigma_n)=0$, the non-zero leading soft behavior ${\cal A}^{(0)_i}_{\rm NLS}(\sigma_n)$ is at the $\tau^1$ order.

\section{Tree SG amplitudes}
\label{sec-SG}

The recursive method in the previous section can also be used to construct the tree SG amplitudes.
The general pure Galileon lagrangian is
\bea
{\cal L}_{\rm Ga}=-{1\over2}\,\partial_\mu\phi\partial^\mu\phi+\sum_{m=3}^\infty\,g_m\,{\cal L}_m\,,~~\label{Lg-SG}
\eea
where
\bea
{\cal L}_m=\phi\,{\rm det}\,\{\partial^{\mu_i}\partial_{\nu_j}\phi\}^{m-1}_{i,j=1}\,.
\eea
The special Galileon theory is the theory with various constraints on coupling constants $g_m$ \cite{Cachazo:2014xea}.
As discussed in subsection.\ref{subsecexpand} in section.\ref{sec-background}, the higher-point vertices can be split into $3$-point ones,
then the consistency requires all created effective coupling constants to have the mass dimension the same as $g_3$, which is $-(2+d)/2$, as can be observed from \eref{Lg-SG}. Then the mass dimension of the kinematic part of $n$-point amplitude should be $(2n-2)$. The lagrangian \eref{Lg-SG} also indicates the absence of decomposition into color ordered partial amplitudes.

Thus, our aim is to construct the un-color-ordered tree amplitudes (kinematic part) for massless scalars, with mass dimension $(2n-2)$. The universality of soft behavior and the double copy structure are also assumed, similar as for the NLSM case. Such amplitudes ${\cal A}_{\rm SG}(\{i\}_n)$ can be expanded to BAS KK basis as in \eref{expan-nocolor}, since the absence of color ordering implies the double summation over $\sigma_{n-2}$ and $\W\sigma_{n-2}$. We add one more assumption that ${\cal C}(\sigma_{n-2},k_i)$
and $\W{\cal C}(\W\sigma_{n-2},k_i)$ are the same function of orderings, i.e.,
\bea
{\cal C}(\sigma_{n-2},k_i)=\W{\cal C}(\W\sigma_{n-2},k_i)\,,~~~~{\rm if}~~ \sigma_{n-2}=\W\sigma_{n-2}\,.~~\label{sym-C}
\eea
In other words, $\W{\cal C}(\W\sigma_{n-2},k_i)$ can be rewritten as ${\cal C}(\W\sigma_{n-2},k_i)$.
This condition fixes the mass dimension of ${\cal C}(\sigma_{n-2},k_i)$
or ${\cal C}(\W\sigma_{n-2},k_i)$ to be $2(n-2)$. We will show that the answer satisfies all above requirements is unique. The construction bears the strong similarity with that in the previous section for obtaining NLSM amplitudes, thus we will omit a variety of details and motivations in this section, since they are already explained. The enhanced Adler zero for SG amplitudes will be discussed at the end of subsection.\ref{subsec-npSG}.

\subsection{$4$-point amplitude and off-shell extension}
\label{subsec-4pSG}

The $3$-point tree SG amplitude has mass dimension $4$, thus does not exist if all external legs are on-shell, due to kinematics. The $4$-point amplitude ${\cal A}_{\rm SG}(\{i\}_4)$ has mass dimension $6$. This amplitude has the symmetry among Mandelstam variables $s$, $t$ and $u$ since it carries no color ordering. The mass dimension and Symmetry ensures that ${\cal A}_{\rm SG}(\{i\}_4)$ is proportional to $s\,t\,u$, thus we can fix it as
\bea
{\cal A}_{\rm SG}(\{i\}_4)=-s\,t\,u\,,~~\label{SG-4p}
\eea
where the overall $-$ sign is introduced for later convenience.

The $4$-point SG amplitude can be expanded to BAS amplitudes as
\bea
{\cal A}_{\rm SG}(\{i\}_4)&=&{\cal C}_1^2\,{\cal A}_{\rm BAS}(1,2,3,4|1,2,3,4)+{\cal C}_2^2\,{\cal A}_{\rm BAS}(1,3,2,4|1,3,2,4)\nn
& &+{\cal C}_1\,{\cal C}_2\,\Big[{\cal A}_{\rm BAS}(1,2,3,4|1,3,2,4)+{\cal A}_{\rm BAS}(1,3,2,4|1,2,3,4)\Big]\,,
\eea
where the assumption of symmetry among ${\cal C}(\sigma_{n-2},k_i)$ and $\W{\cal C}(\W\sigma_{n-2},k_i)$ in \eref{sym-C} is used.
Evaluating BAS amplitudes leads to the equation
\bea
\Big({{\cal C}_1\over s}+{{\cal C}_2\over u}\Big)^2=t^2\,.
\eea
The general solutions without any pole are
\bea
{\cal C}_1=\alpha\,s^2-(1+\beta)\,su\,,~~~~{\cal C}_2=\beta\,u^2-(1+\alpha)\,su\,,~~\label{solu-gener-1}
\eea
and
\bea
{\cal C}_1=\alpha\,s^2+(1-\beta)\,su\,,~~~~{\cal C}_2=\beta\,u^2+(1-\alpha)\,su\,.~~\label{solu-gener-2}
\eea
As discussed around \eref{BCJ1} and \eref{BCJ2}, different values of $\alpha$ and $\beta$ are related by BCJ relations.
We again choose $\alpha=\beta$ to manifest the symmetry among Mandelstam variables $s$ and $u$, then the general solutions in \eref{solu-gener-1}
and \eref{solu-gener-2} are decomposed as the combinations of two special solutions in \eref{solu1-4pNLSM} and \eref{solu2-4pNLSM}.

Similar as for the NLSM case, in order to construct general amplitudes recursively, one need to extend the on-shell amplitudes to the off-shell ones
\bea
A_{\rm SG}(\{i\}_n)=\sum_{\sigma_{n-2}}\,\sum_{\W\sigma_{n-2}}\,C(\sigma_{n-2},k_i)\,C(\W\sigma_{n-2},k_i)\,A_{\rm BAS}(1,\sigma_{n-2},n|1,\W\sigma_{n-2},n)\,,~~\label{expan-SG-offshell}
\eea
which inherit the symmetry between ${\cal C}(\sigma_{n-2},k_i)$ and $\W{\cal C}(\W\sigma_{n-2},k_i)$ in \eref{sym-C}, and do not vanish when the numbers of external particles are odd. The $3$-point off-shell amplitude can be completely fixed by considering the expansion
\bea
A_{\rm SG}(\{i\}_3)&=&C^2\,A_{\rm S}(1,2,3|1,2,3)=(C')^2\,A_{\rm BAS}(3,2,1|3,2,1)\nn
&=&C\,C'\,A_{\rm BAS}(1,2,3|3,2,1)=C\,C'\,A_{\rm BAS}(3,2,1|1,2,3)\,.
\eea
The reversion of the ordering $(1,2,3)\to (3,2,1)$ is accompanied with a $-$ sign, thus $C=-C'$. Since $C$ is associated to the ordering $(1,2,3)$ while $C'$
is associated to $(3,2,1)$, we expect $C$ and $C'$ are related via the replacement $k_1\leftrightarrow k_3$. Two requirements $C=-C'$ and $k_1\leftrightarrow k_3$
lead to the choice $C=2\,k_2\cdot k_1$, $C'=2\,k_2\cdot k_3$ among candidates in \eref{variable}, therefore
\bea
A_{\rm SG}(\{i\}_3)=(2\,k_2\cdot L_2)\,(2\,k_2\cdot R_2)\,A_{\rm BAS}(1,2,3|1,2,3)\,,~~\label{SG-3p-offshell}
\eea
where the combinatory momentum $R_i$ is defined as the summation of momenta at the l.h.s of $i$ in the right ordering carried by the BAS amplitude.

To get the extended $4$-point amplitude, one need to make a particular choice of $C_1$ and $C_2$ among various inequivalent candidates which are equivalent win the on-shell limit. Again, the unique choice can be made by imposing appropriate constraints for poles. For the current SG amplitude without color ordering, we expect that it contain propagators $1/s_{12}$ and $1/s_{13}$ when both $k_1$ and $k_4$ are off-shell. On the other hand, if one of $k_1$ and $k_4$ is assumed to be on-shell, the vanishing of on-shell $3$-point amplitude indicates that the $4$-point amplitude does not contain any pole. The above two requirements for poles completely fix the off-shell coefficients $C_1$ and $C_2$ as
\bea
C_1=(2\,k_1\cdot k_2)\,(2\,k_3\cdot k_4)\,,~~~~C_2=(2\,k_1\cdot k_3)\,(2\,k_2\cdot k_4)\,,
\eea
therefore
\bea
A_{\rm SG}(\{i\}_4)=(2\,k_2\cdot L_2)\,(2\,k_3\cdot L_3)\,(2\,k_2\cdot R_2)\,(2\,k_3\cdot R_3)\,A_{\rm BAS}(1,2\shuffle3,4|1,2\shuffle3,4)\,.~~\label{SG-4p-offshell}
\eea
%

\subsection{Recursive construction for $n$-point amplitude, and enhanced Adler zero}
\label{subsec-npSG}

To construct the general SG amplitudes by applying the recursive method, the first step is to figure out the universal soft behavior. To achieve this goal, we consider the leading order soft behavior of the off-shell $4$-point amplitude in \eref{SG-4p-offshell}. Taking $k_2\to\tau k_2$
and expanding $A_{\rm SG}(\{i\}_4)$ in \eref{SG-4p-offshell} by $\tau$ gives the leading order contribution
\bea
A_{\rm SG}^{(0)_2}(\{i\}_4)&=&\tau\,\Big[(2\,k_2\cdot k_1)^2\,(2\,k_3\cdot k_1)^2+(2\,k_2\cdot k_{13})^2\,(2\,k_3\cdot k_1)^2-2\,(2\,k_2\cdot k_1)\,(2\,k_2\cdot k_{13})\,(2\,k_3\cdot k_1)^2\Big]\nn
& &{1\over s_{23}}\,A_{\rm BAS}(1,3,4|1,3,4)\nn
&=&\tau\,(2\,k_2\cdot k_3)\,A_{\rm SG}(\{i\}_4\setminus2)\,.~~\label{leading1}
\eea
Similarly, considering $k_3\to\tau k_3$ leads to
\bea
A_{\rm SG}^{(0)_3}(\{i\}_4)&=&\tau\,(2\,k_3\cdot k_2)\,A_{\rm SG}(\{i\}_4\setminus3)\,.~~\label{leading2}
\eea
The resulted formulas in \eref{leading1} and \eref{leading2} exhibit the universal soft behavior which can be generalized to the general $n$-point SG amplitude as
\bea
A_{\rm SG}^{(0)_j}(\{i\}_n)&=&S^{(0)_j}_{\rm SG}\,A_{\rm SG}(\{i\}_n\setminus j)\,,~~\label{soft-theo-SG}
\eea
where
\bea
S^{(0)_j}_{\rm SG}=\tau\,\sum_{\ell\neq 1,n,j}\,2\,k_j\cdot k_\ell\,.~~\label{soft-fact-SG}
\eea

Then we can use the recursive method to construct the general off-shell tree SG amplitudes, in the expanded formula. We assume that the $m$-point amplitude can be expanded as
\bea
A_{\rm SG}(\{i\}_m)=\sum_{\sigma_{m-2}}\,\sum_{\W\sigma_{m-2}}\,\Big(\prod_{a=2}^{m-1}\,2\,k_a\cdot L_a\Big)\,\Big(\prod_{b=2}^{m-1}\,2\,k_b\cdot R_b\Big)\,
A_{\rm BAS}(1,\sigma_{m-2},m|1,\W\sigma_{m-2},m)\,.~~\label{SG-mp}
\eea
Obviously, $3$-point and $4$-point amplitudes in \eref{SG-3p-offshell} and \eref{SG-4p-offshell} satisfy this formula. The universal soft behavior in \eref{soft-theo-SG} and \eref{soft-fact-SG} indicates
\bea
A_{\rm SG}^{(0)_j}(\{i\}_{m+1})&=&\tau\,\Big(\sum_{\ell\neq1,m+1,j}\,2k_j\cdot k_\ell\Big)\,A_{\rm SG}(\{i\}_{m+1}\setminus j)\nn
&=&\tau\,\sum_{\sigma_{m-2}}\,\sum_{\W\sigma_{m-2}}\,\Big(\prod_{a\in\{2,\cdots,m\}\setminus j}\,2\,k_a\cdot L_a\Big)\,\Big(\prod_{a\in\{2,\cdots,m\}\setminus j}\,2\,k_b\cdot R_b\Big)\nn
& &\Big(\sum_{\ell\in\{2,\cdots,m\}\setminus j}\,2k_j\cdot k_\ell\Big)\,
A_{\rm BAS}(1,\sigma_{m-2},m+1|1,\W\sigma_{m-2},m+1)\,,~~\label{soft-np-SG-middle}
\eea
where the second equality is obtained by substituting \eref{SG-mp} into the first line.
Here the orderings $\sigma_{m-2}$ and $\W\sigma_{m-2}$ are among external legs in $\{2,\cdots,m\}\setminus j$,
and the combinatory momenta $L_a$ and $R_b$ are defined for left and right orderings $(1,\sigma_{m-2},m+1)$ and $(1,\W\sigma_{m-2},m+1)$
respectively.
To deal with the last line in \eref{soft-np-SG-middle}, we observe that
\bea
& &(k_j\cdot k_\ell)\,A_{\rm BAS}(1,\sigma_{m-2},m+1|1,\W\sigma_{m-2},m+1)\nn
&=&\Big(k_j\cdot (L_j-k_1+X_{(\sigma_{m-2})_j})\Big)\,\delta_{\ell j}\,A_{\rm BAS}(1,\sigma_{m-2}\shuffle {\not j},m+1|1,\W\sigma_{m-2}\shuffle {\not j},m+1)\nn
&=&\Big(k_j\cdot (R_j-k_1+\W X_{(\W\sigma_{m-2})_j})\Big)\,\W\delta_{\ell j}\,A_{\rm BAS}(1,\sigma_{m-2}\shuffle {\not j},m+1|1,\W\sigma_{m-2}\shuffle {\not j},m+1)\,,~~\label{iden-1}
\eea
where $X_{(\sigma_{m-2})_j}$ and $\W X_{(\W\sigma_{m-2})_j}$ are two un-fixed momenta without including $k_j$, $L_j$ and $\delta_{\ell j}$ are defined for the left orderings $(1,\sigma_{m-2}\shuffle j,m+1)$, while $R_j$ and $\W\delta_{\ell j}$ are defined for the right orderings $(1,\W\sigma_{m-2}\shuffle j,m+1)$. Notice that the relation in \eref{iden-1} is correct for any $\ell\in\{2,\cdots,m\}\setminus j$. Substituting \eref{iden-1} into \eref{soft-np-SG-middle},
and using the expression for the leading order soft factor for BAS amplitudes in \eref{soft-fac-s-2}, we obtain
\bea
A_{\rm SG}^{(0)_j}(\{i\}_{m+1})=\sum_{\sigma_{m-1}}\,\sum_{\W\sigma_{m-1}}\,P^j_L(\sigma_{m-1})\,P^j_R(\W\sigma_{m-1})\,
A^{(0)_j}_{\rm BAS}(1,\sigma_{m-1},m+1|1,\W\sigma_{m-1},m+1)\,,~~\label{soft-m+1p-SG}
\eea
where
\bea
P^j_L(\sigma_{m-1})=\Big(\prod_{a\in\{2,\cdots,m\}\setminus j}\,2\,k_a\cdot L^{\not j}_a\Big)\,\Big(k_j\cdot(L_j-k_1+X_{(\sigma_{m-1}\setminus j)_j})\Big)\,,~~\label{PL}
\eea
\bea
P^j_R(\W\sigma_{m-1})=\Big(\prod_{b\in\{2,\cdots,m\}\setminus j}\,2\,k_b\cdot R^{\not j}_b\Big)\,\Big(k_j\cdot(R_j-k_1+\W X_{(\W\sigma_{m-1}\setminus j)_j})\Big)\,.~~\label{PR}
\eea
In \eref{soft-m+1p-SG}, orderings $\sigma_{m-1}$ and $\W\sigma_{m-1}$ are among legs in $\{2,\cdots,m\}$. The combinatory
momenta $L_a$ in \eref{PL} are defined for the left orderings $(1,\sigma_{m-1},m+1)$, while $R_b$ are defined for the right orderings $(1,\W\sigma_{m-1},m+1)$.

The expansion of off-shell SG amplitudes in \eref{expan-SG-offshell} indicates that the leading order conribution can be expressed as
\bea
A_{\rm SG}^{(0)_j}(\{i\}_{m+1})&=&\sum_{\sigma_{m-1}}\,\sum_{\W\sigma_{m-1}}\,C^{(0)_j}(\sigma_{m-1})\, C^{(0)_j}(\W\sigma_{m-1})
\,A^{(0)_j}_{\rm SG}(1,\sigma_{m-1},m+1|1,\W\sigma_{m-1},m+1)\,.~~\label{soft-m+1p-SG2}
\eea
The mass dimension implies the following decompositions
\bea
C(\sigma_{m-1})=\prod_{k=1}^{m-1}\,{\cal K}_k\cdot{\cal K}_{k+m-1}\,,&~~~~& C(\W\sigma_{m-1})=\prod_{k=1}^{m-1}\,\W{\cal K}_k\cdot\W{\cal K}_{k+m-1}\nn
C^{(0)_j}(\sigma_{m-1})=\prod_{k=1}^{m-1}\,{\cal K}^{(0)_j}_k\cdot{\cal K}^{(0)_j}_{k+m-1}\,,&~~~~& C^{(0)_j}(\W\sigma_{m-1})=\prod_{k=1}^{m-1}\,\W{\cal K}^{(0)_j}_k\cdot\W{\cal K}^{(0)_j}_{k+m-1}\,,~~\label{decom}
\eea
where the combinatory momenta ${\cal K}_a$ and $\W{\cal K}_a$ depend on the left and right orderings respectively, ${\cal K}^{(0)_j}_a$ and $\W{\cal K}^{(0)_j}_a$ are generated from ${\cal K}_a$ and $\W{\cal K}_a$ via the rules in \eref{defin-calk}. Substituting \eref{decom} into \eref{soft-m+1p-SG2}, and comparing with \eref{soft-m+1p-SG}, we find
\bea
C(\sigma_{m-1})&=&\prod_{k=1}^{m-1}\,k_j\cdot(L_j-k_1+X_{(\sigma_{m-1}\setminus j)_j})\,,\nn
C(\W\sigma_{m-1})&=&\prod_{k=1}^{m-1}\,k_j\cdot(R_j-k_1+ X_{(\W\sigma_{m-1}\setminus j)_j})\,,
\eea
satisfying
\bea
L^{(0)_\ell}_j-k_1+X^{(0)_{\ell}}_{(\sigma_{m-1}\setminus j)_j}=L^{(0)_\ell}_j\,,~~~~
R^{(0)_\ell}_j-k_1+ X^{(0)_{\ell}}_{(\W\sigma_{m-1}\setminus j)_j}=R^{(0)_\ell}_j\,,~~~~~~{\rm for}~\ell\neq j\,.~~\label{equ-SG}
\eea
Two equations in \eref{equ-SG} uniquely fix $X_{(\sigma_{m-1}\setminus j)_j}$ and $X_{(\W\sigma_{m-1}\setminus j)_j}$
as
\bea
X_{(\sigma_{m-1}\setminus j)_j}= X_{(\W\sigma_{m-1}\setminus j)_j}=k_1\,,
\eea
therefore we arrive at
\bea
A_{\rm SG}(\{i\}_{m+1})=\sum_{\sigma_{m-1}}\,\sum_{\W\sigma_{m-1}}\,\Big(\prod_{a=2}^m\,2\,k_a\cdot L_a\Big)\,\Big(\prod_{b=2}^m\,2\,k_b\cdot R_b\Big)\,
A_{\rm BAS}(1,\sigma_{m-1},m+1|1,\W\sigma_{m-1},m+1)\,.~~\label{SG-m+1p-final}
\eea
Expressions in \eref{SG-mp} and \eref{SG-m+1p-final} lead to the expanded formula of general off-shell SG amplitudes
\bea
A_{\rm SG}(\{i\}_n)=\sum_{\sigma_{n-2}}\,\sum_{\W\sigma_{n-2}}\,\Big(\prod_{a=2}^{n-1}\,2\,k_a\cdot L_a\Big)\,\Big(\prod_{b=2}^{n-1}\,2\,k_b\cdot R_b\Big)\,
A_{\rm BAS}(1,\sigma_{n-2},n|1,\W\sigma_{n-2},n)\,.~~\label{SG-np}
\eea

The on-shell physical amplitudes are obtained from \eref{SG-np} by setting $k_1^2=k_n^2=0$. The amplitudes with odd number of external legs vanish, while those with even number of legs are expanded as
\bea
{\cal A}_{\rm SG}(\{i\}_n)=\sum_{\sigma_{n-2}}\,\sum_{\W\sigma_{n-2}}\,\Big(\prod_{a=2}^{n-1}\,2\,k_a\cdot L_a\Big)\,\Big(\prod_{b=2}^{n-1}\,2\,k_b\cdot R_b\Big)\,
{\cal A}_{\rm BAS}(1,\sigma_{n-2},n|1,\W\sigma_{n-2},n)\,,~~\label{onshell-SG-np}
\eea
coincide with the result found in \cite{Zhou:2019mbe}.

The enhanced Adler zero for SG amplitudes is more interesting than that for NLSM amplitudes. The naive power counting for \eref{onshell-SG-np} shows that ${\cal A}^{(0)_i}_{\rm SG}(\{i\}_n)$ is at the $\tau^{1}$ order. Count in a more careful way, one can sum over $\W\sigma_{n-2}$ (or $\sigma_{n-2}$) in \eref{onshell-SG-np} to get
\bea
{\cal A}_{\rm SG}(\{i\}_n)=\sum_{\sigma_{n-2}}\,\Big(\prod_{a=2}^{n-1}\,2\,k_a\cdot L_a\Big)\,
{\cal A}_{\rm NMS}(1,\sigma_{n-2},n)\,,
\eea
by comparing \eref{onshell-SG-np} with \eref{expan-onshellNLSM-np}.
Since the leading order of $\prod_{a=2}^{n-1}\,(2\,k_a\cdot L_a)^{(0)_i}$ is the $\tau^1$ order, while the leading order of ${\cal A}^{(0)_i}_{\rm NMS}(1,\sigma_{n-2},n)$ is also the $\tau^1$ order, one may conclude that ${\cal A}^{(0)_i}_{\rm SG}(\{i\}_n)$ is at the $\tau^2$ order. From the top down perspective, counting of derivatives also leads to the $\tau^2$ order.
However, different from above results, ${\cal A}^{(0)_i}_{\rm SG}(\{i\}_n)$ is indeed at the $\tau^3$ order, as can be seen in the following interpretation.

Using the leading order soft theorem for BAS amplitudes in \eref{soft-theo-s} and \eref{soft-fac-s-0}, one can obtain that
the soft behavior of off-shell SG amplitudes in \eref{soft-theo-SG} and \eref{soft-fact-SG} is modified to
\bea
{\cal A}^{[\tau^1]_j}_{\rm SG}(\{i\}_n)=\tau\,\Big(\sum_{\ell\neq j}\,2\,k_j\cdot k_{\ell}\Big)\,\Big[\lim_{k_1^2\to0,k_2^2\to0}\,A_{\rm SG}(\{i\}_n\setminus j)\Big]\,,
\eea
where the superscript $[\tau^1]_j$ encodes the $\tau^1$ order contribution when $k_j\to\tau k_j$.
The factor $\sum_{\ell\neq j}\,k_j\cdot k_{\ell}$ vanishes due to the momentum conservation and the on-shell condition, while $\lim_{k_1^2\to0,k_2^2\to0}\,A_{\rm SG}(\{i\}_n\setminus j)$ vanishes due to the odd number of external legs. This is the soft behavior at the $\tau^1$ order.

To discuss the soft behavior at the $\tau^2$ order, we first sum over $\sigma_{n-2}$ or $\W\sigma_{n-2}$ to rewrite ${\cal A}_{\rm SG}(\{i\}_n)$ as
\bea
{\cal A}_{\rm SG}(\{i\}_n)&=&\sum_{\sigma_{n-2}}\,\sum_{\W\sigma_{n-2}}\,{\cal C}(\sigma_{n-2},k_i)\,{\cal C}(\W\sigma_{n-2},k_i)\,
{\cal A}_{\rm BAS}(1,\sigma_{n-2},n|1,\W\sigma_{n-2},n)\nn
&=&\sum_{\sigma_{n-2}}\,{\cal C}(\sigma_{n-2},k_i)\,
{\cal A}_{\rm NLS}(1,\sigma_{n-2},n)\nn
&=&\sum_{\W\sigma_{n-2}}\,{\cal C}(\W\sigma_{n-2},k_i)\,
{\cal A}_{\rm NLS}(1,\W\sigma_{n-2},n)\,.
\eea
Then, we split the $\tau^2$ contributions as
\bea
{\cal A}^{[\tau^2]_j}_{\rm SG}(\{i\}_n)=B_1+B_2+B_3\,,
\eea
where
\bea
B_1&=&\sum_{\sigma_{n-2}}\,\sum_{\W\sigma_{n-2}}\,{\cal C}^{(1)_j}(\sigma_{n-2},k_i)\,{\cal C}^{(0)_j}(\W\sigma_{n-2},k_i)\,
{\cal A}^{(0)_j}_{\rm BAS}(1,\sigma_{n-2},n|1,\W\sigma_{n-2},n)\nn
&=&\sum_{\sigma_{n-2}}\,{\cal C}^{(1)_j}(\sigma_{n-2},k_i)\,
{\cal A}^{[\tau^0]_j}_{\rm NLS}(1,\sigma_{n-2},n)\,,\nn
B_2&=&\sum_{\sigma_{n-2}}\,\sum_{\W\sigma_{n-2}}\,{\cal C}^{(0)_j}(\sigma_{n-2},k_i)\,{\cal C}^{(1)_j}(\W\sigma_{n-2},k_i)\,
{\cal A}^{(0)_j}_{\rm BAS}(1,\sigma_{n-2},n|1,\W\sigma_{n-2},n)\nn
&=&\sum_{\W\sigma_{n-2}}\,{\cal C}^{(1)_j}(\W\sigma_{n-2},k_i)\,
{\cal A}^{[\tau^0]_j}_{\rm NLS}(1,\W\sigma_{n-2},n)\,,
\eea
while
\bea
B_3&=&\sum_{\sigma_{n-2}}\,\sum_{\W\sigma_{n-2}}\,{\cal C}^{(0)_j}(\sigma_{n-2},k_i)\,{\cal C}^{(0)_j}(\W\sigma_{n-2},k_i)\,
{\cal A}^{(1)_j}_{\rm BAS}(1,\sigma_{n-2},n|1,\W\sigma_{n-2},n)\,.
\eea
Here the superscript $(1)_j$ denotes the contributions at the sub-leading order when considering the soft behavior for the external momentum $k_j$. Two blocks $B_1$ and $B_2$ vanish due to
\bea
{\cal A}^{[\tau^0]_j}_{\rm NLS}(1,\sigma_{n-2},n)={\cal A}^{[\tau^0]_j}_{\rm NLS}(1,\W\sigma_{n-2},n)=0\,,
\eea
which describe the enhanced Adler zero of NLSM amplitudes discussed in the previous section. The block $B_3$ can be rewritten as
\bea
B_3&=&\tau^2\,\sum_{\sigma_{n-3}}\,\sum_{\W\sigma_{n-3}}\,{\cal C}^{(0)_j}(\sigma_{n-3},k_i)\,{\cal C}^{(0)_j}(\W\sigma_{n-3},k_i)\nn
& &\Big[
(2\,k_j\cdot L_j)\,(2\,k_j\cdot R_j)\,{\cal A}^{(1)_j}_{\rm BAS}(1,\sigma_{n-3}\shuffle j,n|1,\W\sigma_{n-3}\shuffle j,n)\Big]\,,
\eea
where orderings $\sigma_{n-3}$ and $\W\sigma_{n-3}$ are among legs in $\{2,\cdots,n-1\}\setminus j$. It is straightforward to observe that two coefficients ${\cal C}^{(0)_j}(\sigma_{n-3},k_i)$ and ${\cal C}^{(0)_j}(\W\sigma_{n-3},k_i)$ are independent of the external momentum $k_j$ and the positions of $j$ in two orderings, thus $B_3$ vanishes due to the fundamental BCJ relation \eref{bcj}. Consequently, the soft behavior of SG amplitudes also vanish at the $\tau^2$ order.

\section{Enhanced Adler zero for BI and DBI}
\label{sec-expan-toNLSM}

In this section, we briefly discuss the enhanced Adler zero for tree amplitudes of Born-Infeld (BI) theory, as well as the scalar sector of tree amplitudes of Dirac-Born-Infeld (DBI) theory. We will not construct those amplitudes recursively as in previous subsections, based on the following reasons. For BI amplitudes, one of two coefficients ${\cal C}(\sigma_{n-2},k_i)$ and $\W{\cal C}(\W\sigma_{n-2},k_i)$ in \eref{expan-nocolor} should depend on the polarizations carried by BI photons, thus the lowest-point amplitude can not be fixed by considering only Mandelstam variables. This is an interesting challenge, and we leave it to the future work. The scalar sector of tree DBI amplitudes are those all external legs are scalars. For these DBI amplitudes, the method used in section.\ref{sec-expan} and section.\ref{sec-SG} is not useful, due to the following reason. For NLSM amplitudes, the coefficient ${\cal C}(\sigma_{n-2},k_i)$ in \eref{exp-N-KK} has mass dimension $2(n-2)$, and for SG amplitudes, the total mass dimension of two coefficients ${\cal C}(\sigma_{n-2},k_i)$ and $\W{\cal C}(\W\sigma_{n-2},k_i)$ is $4(n-2)$. These numbers are even when $n$ is odd, it means one can always construct the Lorentz invariant coefficients from Mandelstam variables, thus the definition for off-shell amplitudes with odd numbers of external legs is allowed. However, the total mass dimension of ${\cal C}(\sigma_{n-2},k_i)$ and $\W{\cal C}(\W\sigma_{n-2},k_i)$ for DBI amplitudes is turned to be $3(n-2)$, which becomes odd when $n$ is odd. Thus, for odd $n$, it is impossible to construct any Lorentz invariant with correct mass dimension, by using only external momenta. This observation indicates that the well-defined off-shell DBI amplitude with odd number of external legs does not exist, thus the previous recursive method fall to work.

On the other hand, the coefficients ${\cal C}(\sigma_{n-2},k_i)$ and $\W{\cal C}(\W\sigma_{n-2},k_i)$ for BI and DBI amplitudes can be determined through other methods, such as deriving them from CHY formalism, or constructing them via differential operators. When the correct ${\cal C}(\sigma_{n-2},k_i)$ and $\W{\cal C}(\W\sigma_{n-2},k_i)$ are given, one can discuss the enhanced Adler zero for these amplitudes. This is the task of the current section.

\subsection{Enhance Adler zero for BI}
\label{subsec-BI}

For clarity, let us denote coefficients ${\cal C}(\sigma_{n-2},k_i)$ in \eref{exp-N-KK} and \eref{expan-nocolor} by ${\cal C}_{\rm NLS}(\sigma_{n-2},k_i)$. Then, the SG amplitudes are those with
\bea
{\cal C}(\sigma_{n-2},k_i)={\cal C}_{\rm NLS}(\sigma_{n-2},k_i)\,,~~~~\W{\cal C}(\W\sigma_{n-2},k_i)={\cal C}_{\rm NLS}(\W\sigma_{n-2},k_i)\,.
\eea
Other amplitudes can be generated by replacing ${\cal C}(\sigma_{n-2},k_i)$ or $\W{\cal C}(\W\sigma_{n-2},k_i)$ by different appropriate candidates.
The BI amplitudes are those with
\bea
{\cal C}(\sigma_{n-2},k_i)={\cal C}_{\rm NLS}(\sigma_{n-2},k_i)\,,~~~~\W{\cal C}(\W\sigma_{n-2},k_i)={\cal C}_{\rm YM}(\W\sigma_{n-2},k_i,\epsilon_i)\,,
\eea
namely,
\bea
{\cal A}_{\rm BI}(\{i\}_n)=\sum_{\sigma_{n-2}}\,\sum_{\W\sigma_{n-2}}\,{\cal C}_{\rm NLS}(\sigma_{n-2},k_i)\,{\cal C}_{\rm YM}(\W\sigma_{n-2},k_i,\epsilon_i)\,{\cal A}_{\rm BAS}(1,\sigma_{n-2},n|1,\W\sigma_{n-2},n)\,,~~\label{expan-BI-KK}
\eea
see in \cite{Cachazo:2014xea,Zhou:2019mbe}. Here ${\cal C}_{\rm YM}(\W\sigma_{n-2},k_i,\epsilon_i)$ serve as the coefficients when expanding color ordered pure Yang-Mills (YM) amplitudes
${\cal A}_{\rm YM}(\sigma_n)$ to BAS KK basis as
\bea
{\cal A}_{\rm YM}(\sigma_n)=\sum_{\W\sigma_{n-2}}\,{\cal C}_{\rm YM}(\W\sigma_{n-2},k_i,\epsilon_i)\,{\cal A}_{\rm BAS}(\sigma_n|1,\W\sigma_{n-2},n)\,.
\eea
Summing over $\W\sigma_{n-2}$ in \eref{expan-BI-KK} gives the expansion of BI amplitudes to YM ones
\bea
{\cal A}_{\rm BI}(\{i\}_n)=\sum_{\sigma_{n-2}}\,\Big(\prod_{\ell=2}^{n-1}\,2k_\ell\cdot L_\ell\Big)\,{\cal A}_{\rm YM}(1,\sigma_{n-2},n)\,.~~\label{expan-BI-toYM}
\eea

The naive power counting for \eref{expan-BI-toYM} shows that ${\cal A}^{(0)_i}_{\rm BI}(\{i\}_n)$ is at the $\tau^0$ order, since $\tau$ arise from coefficients cancel $\tau^{-1}$ from YM basis. However, the soft behavior indeed vanish at the $\tau^0$ order, exhibits the enhanced Adler zero from the bottom up perspective.
The numbers of external legs of un-vanishing on-shell BI amplitudes are also even. However, one can consider the extended off-shell BI amplitudes
\bea
A_{\rm BI}(\{i\}_n)=\sum_{\sigma_{n-2}}\,\Big(\prod_{\ell=2}^{n-1}\,2k_\ell\cdot L_\ell\Big)\,A_{\rm YM}(1,\sigma_{n-2},n)\,,~~\label{expan-BI-toYM-offshell}
\eea
with $k_1^2\neq0$, $k_2^2\neq0$, which do not vanish when $n$ is odd.
The leading order single soft behavior of such off-shell BI amplitudes $A_{\rm BI}(\{i\}_n)$ can be figured out by using the expansion \eref{expan-BI-toYM}
and the soft behavior of off-shell YM amplitudes $A_{\rm YM}(1,\sigma_{n-2},n)$.
For the on-shell YM amplitudes ${\cal A}_{\rm YM}(\sigma_n)$, the leading order soft theorem for the single external gluon $i$ is given as
\bea
{\cal A}^{(0)_i}_{\rm YM}(\sigma_n)=S^{(0)_i}_{\rm YM}\,{\cal A}_{\rm YM}(\sigma_n\setminus i)\,,
\eea
where the universal soft factor is
\bea
S^{(0)_i}_{\rm YM}={1\over\tau}\,\sum_{j\neq i}\,{\delta_{ji}\,(\epsilon_i\cdot k_j)\over s_{ji}}\,.~~\label{soft-fac-gluon}
\eea
For the off-shell YM amplitudes ${\cal A}^{(0)_i}_{\rm YM}(\sigma_n)$, the leading order soft factor is modified by eliminating
$j=1$ and $j=n$ in the summation in \eref{soft-fac-gluon}, since $s_{1i}$ and $s_{ni}$ are no longer proportional to $\tau$.
With the soft theorem provided above, one can decompose the leading order contribution of the BI amplitude as
\bea
A^{(0)_j}_{\rm BI}(\{i\}_n)&=&\tau\,\sum_{\sigma_{n-2}}\,\Big(2\,k_j\cdot L_j\Big)\,\Big(\prod_{\ell\in\{2,\cdots,n-1\}\setminus j}\,2k_\ell\cdot L^{\not j}_\ell\Big)\,A^{(0)_j}_{\rm YM}(1,\sigma_{n-2},n)\nn
&=&\sum_{\sigma_{n-2}}\,\Big(\prod_{\ell\in\{2,\cdots,n-1\}\setminus j}\,2k_\ell\cdot L^{\not j}_\ell\Big)\,\Big(2\,k_j\cdot L_j\Big)\,\Big(\sum_{a\in\{2,\cdots,n-1\}\setminus j}\,{\delta_{aj}\,(\epsilon_j\cdot k_a)\over s_{aj}}\Big)\,A_{\rm YM}(1,\sigma_{n-2}\setminus j,n)\nn
&=&\Big(\sum_{a\in\{2,\cdots,n-1\}\setminus j}\,\epsilon_j\cdot k_a\Big)\,\Big[\sum_{\sigma_{n-3}}\,\Big(\prod_{\ell\in\{2,\cdots,n-1\}\setminus j}\,2k_\ell\cdot L_\ell\Big)\,A_{\rm YM}(1,\sigma_{n-3},n)\Big]\nn
&=&\Big(\sum_{a\in\{2,\cdots,n-1\}\setminus j}\,\epsilon_j\cdot k_a\Big)\,A_{\rm BI}(\{i\}_n\setminus j)\,.~~\label{soft-BI}
\eea
In the first and second lines, $L_j$ and $L_{\ell}$ are defined for the ordering $(1,\sigma_{n-2},n)$, while in the third line $L_{\ell}$ are defined for the ordering $(1,\sigma_{n-3},n)$, where $\sigma_{n-3}=\sigma_{n-2}\setminus j$. The third line is derived by interchanging the order of two summations in the second line. The last equality uses the expansion \eref{expan-BI-toYM} for the $(n-1)$-point amplitude. Here $\sum_{a\in\{2,\cdots,n-1\}\setminus j}\,\epsilon_j\cdot k_a$ serves as the universal soft factor.

Back to the physical on-shell case, we see that $j=1$ and $j=n$ return to the summation in \eref{soft-fac-gluon}, thus the soft factor is modified to
$\sum_{a\neq j}\,\epsilon_j\cdot k_a$. This new soft factor vanishes because of the momentum conservation and on-shell condition $\epsilon_j\cdot k_j=0$. The amplitude $\lim_{k_1^2\to0,k_2^2\to0}\,A_{\rm BI}(\{i\}_n\setminus j)$ vanishes simultaneously, since the odd number of external legs. This is the interpretation for the enhanced Adler zero for BI amplitudes.

\subsection{Enhanced Adler zero for DBI}
\label{subsec-expan-NLSM}

The enhanced Adler zero for BI amplitudes can also be understood in another way. Summing over $\sigma_{n-2}$ in \eref{expan-BI-KK} instead of $\W\sigma_{n-2}$, one get
\bea
{\cal A}_{\rm BI}(\{i\}_n)=\sum_{\W\sigma_{n-2}}\,{\cal C}_{\rm YM}(\W\sigma_{n-2},k_i,\epsilon_i)\,{\cal A}_{\rm NLS}(1,\W\sigma_{n-2},n)\,,
\eea
which is the expansion of BI amplitudes to NLSM ones.
Through this path, it is complicated to obtain the elegant soft behavior in \eref{soft-BI}, but is sufficient to observe the vanishing of ${\cal A}^{[\tau^0]_j}_{\rm BI}(\{i\}_n)$ from the vanishing of ${\cal A}^{[\tau^0]_j}_{\rm NLS}(1,\W\sigma_{n-2},n)$.

The enhanced Adler zero for the scalar sector of DBI amplitudes can be understood similarly. The DBI amplitudes are those with
\bea
{\cal C}(\sigma_{n-2},k_i)={\cal C}_{\rm NLS}(\sigma_{n-2},k_i)\,,~~~~\W{\cal C}(\W\sigma_{n-2},k_i)={\cal C}_{\rm sYS}(\W\sigma_{n-2},k_i)\,,
\eea
where ${\cal C}_{\rm sYS}(\W\sigma_{n-2},k_i)$ are coefficients in the expansion of the color ordered special Yang-Mills-scalar amplitudes ${\cal A}_{\rm sYS}(\sigma_n)$ to BAS KK basis. Here the special Yang-Mills-scalar theory is the low energy effective theory of $N$ coincident $D$-branes \cite{Cachazo:2014xea}.
Summing over $\sigma_{n-2}$ leads to
\bea
{\cal A}_{\rm DBI}(\{i\}_n)=\sum_{\W\sigma_{n-2}}\,{\cal C}_{\rm sYS}(\W\sigma_{n-2},k_i)\,{\cal A}_{\rm NLS}(1,\W\sigma_{n-2},n)\,,
\eea
therefore
\bea
{\cal A}^{(0)_j}_{\rm DBI}(\{i\}_n)=\sum_{\W\sigma_{n-2}}\,{\cal C}^{(0)_j}_{\rm sYS}(\W\sigma_{n-2},k_i)\,{\cal A}^{(0)_j}_{\rm NLS}(1,\W\sigma_{n-2},n)\,.
\eea
The coefficients ${\cal C}_{\rm sYS}(\W\sigma_{n-2},k_i)$ do not have the formula as compact as ${\cal C}_{\rm NLS}(\W\sigma_{n-2},k_i)$,
but can be evaluated via the systematic rules in \cite{Zhou:2019mbe}. Using these rules, one can observe that ${\cal C}^{(0)_j}_{\rm sYS}(\W\sigma_{n-2},k_i)$
is at the $\tau^1$ order. On the other hand, ${\cal A}^{(0)_j}_{\rm NLS}(1,\W\sigma_{n-2},n)$ is also at the $\tau^1$ order. Consequently, the leading soft behavior ${\cal A}^{(0)_j}_{\rm DBI}(\{i\}_n)$ is at the $\tau^2$ order.

\section{Summery}
\label{sec-conclu}

In this paper, we proposed a new bottom up method to construct tree NLSM and SG amplitudes recursively. The construction is based on two assumptions which are the universality of soft behaviors, and the double copy structure. To realize the recursive pattern, the off-shell extension of amplitudes which allows the numbers of external legs to be odd are also introduced. Then we bootstrapped the $3$-point and $4$-point amplitudes, and derived the soft behaviors of $4$-point NLSM and SG amplitudes from them. Due to the universality of soft behaviors, we inverted the resulted soft theorems to construct higher-point off-shell amplitudes. Together with some natural requirements such as appropriate mass dimension and permutation invariance among legs in $\{2,\cdots,n-1\}$, the general off-shell amplitudes with arbitrary number of external legs were uniquely determined, expressed in the formula of expanding them to KK BAS basis. Since the universal soft behaviors are derived rather than assumed, the whole procedure is self-contained. In the on-shell limit, amplitudes with odd numbers of external legs vanish automatically. The on-shell limit also gives the new understanding of enhanced Adler zero. From the bottom up point of view, the enhanced Adler zero was described as that amplitudes have vanished soft behavior which exceed the degree expected from the naive power counting of soft momentum in the expanded formula. An interesting observation is that such "zero" have explicit formulas. For example, the enhanced Adler zero for NLSM amplitudes can be interpreted by the simple identity \eref{iden} and the vanishing of amplitudes with odd number of external legs. For tree BI and DBI amplitudes, the current version of construction does not make sense, however the enhanced Adler zero of these amplitudes can be understood similarly.

Through the process of construction, we have not explicitly exploited the factorization near associated poles. However, since the leading soft behavior arises as the factorization at poles associated with $2$-point channels, inverting soft theorem is equivalent to gluing $3$-point off-shell amplitudes to the original amplitude via a special way. Thus the unitarity and locality are preserved automatically. From this point of view, the universality of soft behavior is ensured by the uniqueness of lowest $3$-point amplitudes, thus it is natural to expect that the universality of soft behavior can be satisfied by a variety of other theories.

Various generalizations of the current method can be considered, including applications to other theories and the loop level. In the recent work in \cite{Brown:2023srz}, the BCJ relations among color ordered amplitudes, together with unitarity and locality, were used to bootstrap color ordered tree amplitudes of pure scalar effective theories. In our construction, the ansatz in \eref{exp-N-KK} (with subscript "NLS" replaced by other theories) indicates that the constraints from BCJ relations are satisfied automatically, and the constructions for amplitudes of different theories are shifted to constructions for different coefficients. The advantage of employing the ansatz in \eref{exp-N-KK} is that one need not to solve equations imposed by BCJ relations. The ansatz in \eref{exp-N-KK} still focus on amplitudes which satisfy the double copy structure. Perhaps the most interesting generalization of our method is applying to amplitudes without double copy structure, since a wide range of amplitudes do not have such luxuries.

\section*{Acknowledgments}

The author would thank Prof. Yijian Du and Prof. Song He for useful discussions and suggestions.


\end{document}